\documentstyle[12pt,aaspp4]{article}
\def	\beq	{\begin{equation}}
\def	\eeq	{\end{equation}}

\def	\Angstrom	{\,{\rm \AA}}		% Angstrom

\def	\H	{{\rm H}}
\def	\HH	{{\rm H}_2}
\def	\He	{{\rm He}}
\def	\K	{{\rm K}}
\def	\Re	{{\rm Re}}
\def	\ahat	{\hat{\bf a}}

\def	\bB	{{\bf B}}

\def 	\bE	{{\bf E}}
\def	\bF	{{\bf F}}
\def	\bGam	{{\bf \Gamma}}

\def	\bI	{{\bf I}}
\def	\bJ	{{\bf J}}
\def	\bM	{{\bf M}}
\def	\bQ	{{\bf Q}}

\def	\bV	{{\bf V}}

\def	\be	{{\bf e}}
\def	\bk	{{\bf k}}

\def	\bp	{{\bf p}}
\def	\br	{{\bf r}}
\def	\cm	{\,{\rm cm}}
\def	\DG	{{\rm DG}}
\def	\eff	{{\rm eff}}
\def	\ehat	{\hat{\bf e}}
\def	\erg	{\,{\rm ergs}}
\def	\eV	{\,{\rm eV}\,}
\def	\g	{\,{\rm g}}
\def	\iso	{{\rm iso}}
\def	\ISRF	{{\rm ISRF}}
\def	\khat	{\hat{\bf k}}
\def	\lambdabar	{\bar{\lambda}}
\def	\micron	{\mu{\rm m}}
\def	\nH	{n_{\rm H}}
\def	\nhat	{\hat{\bf n}}

\def	\s	{\,{\rm s}}
\def    \simlt  {\lower.5ex\hbox{$\; \buildrel < \over \sim \;$}}
\def    \simgt  {\lower.5ex\hbox{$\; \buildrel > \over \sim \;$}}
\def	\gtsim	{\simgt}
\def	\ltsim	{\simlt}
\def	\yr	{{\rm yr}}

\def	\nature	{{\it Nature}}
\def	\science {{\it Science}}

\lefthead{Draine \& Weingartner}
\righthead{Radiative Torques on Interstellar Grains}

\begin{document}

\author{{\it Astrophysical Journal}, {\bf470}, in press (Oct. 10)}
\title{Radiative Torques on Interstellar Grains:\\
	I. Superthermal Spinup}
\author{B.T. Draine}
\affil{Princeton University Observatory, Peyton Hall,
	Princeton, NJ 08544, USA; draine@astro.princeton.edu}

\and

\author{Joseph C. Weingartner}
\affil{Physics Dept., Jadwin Hall, Princeton University,
	Princeton, NJ 08544, USA; josephw@phoenix.princeton.edu}

\begin{abstract}
Irregular dust grains are subject to radiative torques when irradiated
by interstellar starlight.
It is shown how these radiative torques may be calculated using the 
discrete dipole approximation.
Calculations are carried out for one irregular grain
geometry, and three different grain sizes.
It is shown that radiative torques can play an important
dynamical role in spinup of interstellar dust grains, 
resulting in rotation rates which may exceed even those expected
from $\HH$ formation on the grain surface.
Because the radiative torque on an interstellar grain is
determined by the overall grain
geometry rather than merely the condition of the grain surface,
the resulting superthermal rotation
is expected to be quite long-lived.
By itself, long-lived superthermal rotation would permit grain alignment
by normal paramagnetic dissipation on the ``Davis-Greenstein''
timescale $\tau_\DG$.
However, radiative torques arising from anisotropy of the starlight
background can act directly to alter the grain
alignment on times short compared to $\tau_\DG$.
Radiative torques must therefore
play a central role in the process of interstellar grain alignment.

The radiative torques depend strongly on the grain size, measured by
$a_\eff$, the radius of a sphere of equal volume.
In diffuse clouds,
radiative torques dominate the torques due to $\HH$ formation for
$a_\eff=0.2\micron$ grains, but are relatively
unimportant for $a_\eff\leq0.05\micron$ grains. 
We argue that this may
provide a natural explanation for the observation
that $a_\eff\gtsim0.1\micron$ grains in diffuse clouds are aligned, while 
there is very little alignment of $a_\eff\ltsim0.05\micron$ grains.
We show that radiative torques are ineffective at producing superthermal
rotation within quiescent dark clouds,
but can be very effective in star-forming regions such as the M17 molecular
cloud.

\end{abstract}


\section{Introduction}

Polarization of starlight by aligned interstellar dust grains was
discovered nearly half a century ago
(Hiltner 1949a,b; Hall 1949; Hall \& Mikesell 1949), but the processes responsible for the
observed alignment remain uncertain.
Davis \& Greenstein (1951) observed that interstellar grains were expected to
be rotating rapidly due to Brownian motion, and proposed that these spinning
grains could be aligned with the local magnetic field by paramagnetic
dissipation.
However,
further study of the statistical mechanics of grain alignment
(Jones \& Spitzer 1967; Purcell \& Spitzer 1971) raised
questions about the ability of the interstellar magnetic field
to achieve the observed
degree of alignment,
since random gas-grain collisions would tend to
oppose the alignment process.

Purcell (1975, 1979) first recognized that interstellar grains were expected 
to have
superthermal rotational velocities as the result of systematic torques,
the most important of which appeared to be due to the process
of $\HH$ formation on the grain surface.
As discussed by Purcell,
superthermal rotation due to torques which are fixed in body coordinates 
can enhance the degree of
grain alignment, since the ``thermal'' torques due to
random collisions with gas atoms now have little effect on the direction of
the grain
angular momentum, allowing paramagnetic dissipation to inexorably bring the
angular momentum into alignment with the galactic field.

The systematic torques considered by Purcell were due
to processes taking place at the grain surface -- $\HH$ formation,
photoelectric emission, and inelastic collisions with gas atoms -- 
but the grain surface
may be altered due to contamination or erosion on relatively
short time scales.
The resulting changes in direction of the systematic torque can cause the
grain to occasionally undergo short periods when its rotation is
``spun down''; during these ``crossover''
episodes the grain may become disaligned (Spitzer \& McGlynn 1979).
Because of this disorientation during ``crossover'', it was not clear 
whether ordinary
paramagnetic dissipation plus superthermal rotation driven by
``Purcell torques'' can account for
the observed grain alignment.
%BTD 96.04.16 This led to
In addition, dust grains were observed to be aligned in some dense molecular
regions where ``Purcell torques'' were expected to be ineffective due to
a low H/H$_2$ ratio, attenuation of the ultraviolet radiation required
for photoelectric emission, and near-equality of gas and grain temperatures.
As a result, there has been
renewed interest in alternatives, including the
possibility that grains may be superparamagnetic (Jones \& Spitzer 1967;
Duley 1978; Martin 1995; Goodman \& Whittet 1995), 
or that grain alignment is due to gas-grain
streaming (Gold 1952; Lazarian 1994, 1995a; Roberge, Hanany, \& Messinger 1995).
Lazarian (1995b) has emphasized the possible importance of grain
``helicity'', since helical grains can be driven to superthermal rotational
velocities, and possibly aligned, when exposed to either streaming gas
atoms or anisotropic radiation.

Harwit (1970a,b) suggested that the quantized angular momentum of the photon
could lead to rapid rotation of a grain following absorption and emission
of many photons, and proposed that the anisotropy of starlight could
result in a tendency for interstellar grains to spin with their angular
momentum 
%BTD 96.04.16 axes 
vectors 
parallel to the Galactic plane.
Dolginov (1972) observed that interstellar grains might have different
absorption and scattering cross sections for left- and right-handed
circularly polarized light, so that the grain angular
momentum could be changed if illuminated by unpolarized but anisotropic
radiation.
This effect was further discussed by Dolginov \& Mytrophanov (1976)
for particles in the Rayleigh limit,
and by Dolginov \& Silant'ev (1976) for larger particles but with refractive
index close to unity so that the Rayleigh-Gans approximation could be used.
Dolginov \& Mytrophanov noted that this process could lead to both 
rapid rotation and possible alignment.  
Unfortunately, Dolginov and collaborators were
unable to calculate the torques on grains with realistic compositions and
sizes.

\begin{figure}
\epsscale{1.00}
\plotone{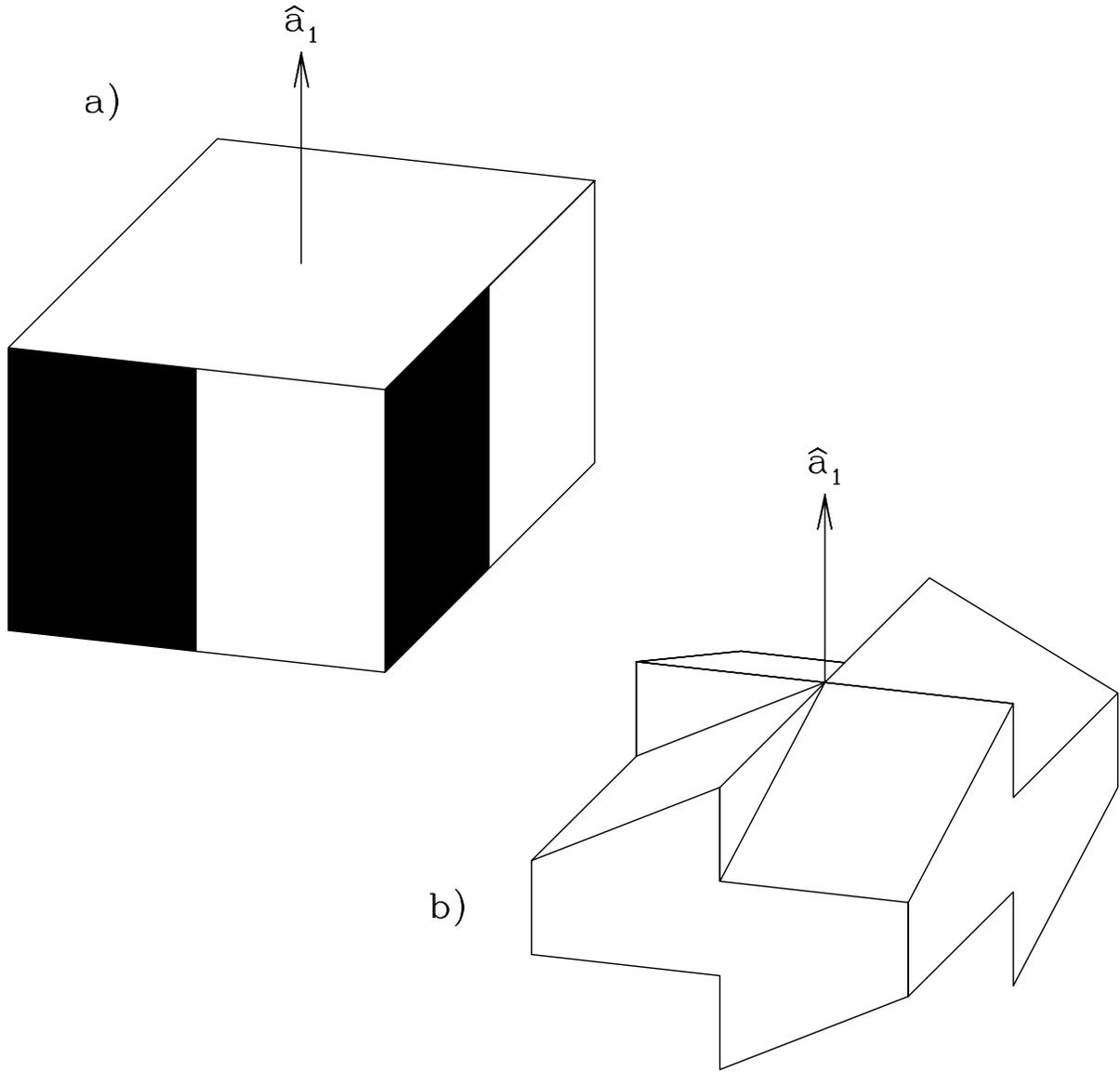}
\caption{
	\label{fig:targs}
	Two examples of macroscopic targets.
	(a) A target on which
	an isotropic radiation field would exert a torque along
	axis $\ahat_1$.  The target is perfectly reflecting except
	for the regions painted black.
	(b) A perfectly-reflecting target on which an isotropic 
	radiation field would exert zero torque, 
	radiation antiparallel to $\ahat_1$ would exert a positive
	torque parallel to $\ahat_1$,
	and radiation parallel to $\ahat_1$ would exert a torque
	antiparallel to $\ahat_1$.
	}
\end{figure}

That irregular interstellar grains should be subject to radiative torques
is not surprising.
We consider two macroscopic examples for illustration.
Fig.\ \ref{fig:targs}a shows a four-fold symmetric target with square
top and bottom, with 
each of its four rectangular
sides divided into perfectly absorbing and perfectly
reflecting halves.
In the geometric optics limit the normal component of the radiation pressure
force will be twice as large on the reflecting sections as on the
absorbing sections; as a result, an isotropic radiation field illuminating
this target will produce
a positive torque along axis
$\ahat_1$.
However, incident radiation which is either parallel or antiparallel to
$\ahat_1$ will not produce any torque
%BTD 96.04.16
on this target.

Fig.\ \ref{fig:targs}b shows a target obtained
by starting with the shape of Fig.\ref{fig:targs}a and removing four wedges
from the top and four from the bottom, with the resulting shape being
symmetric under reflection through the centroid.
Fig.\ref{fig:targs}b is an example of a shape with ``helicity''.
From symmetry it is clear that an isotropic radiation
field will produce no torque on this target: whatever torque is exerted on
the ``top'' half of the grain will be cancelled by an opposite torque on
the ``bottom'' half.
However, anisotropic illumination can produce a torque on this target.
If, for example, the surfaces are all perfectly reflecting, then 
incident radiation which is antiparallel to $\ahat_1$ will produce a torque
parallel to $\ahat_1$, while radiation parallel to $\ahat_1$ will produce
a torque antiparallel to $\ahat_1$.
These two examples show that macroscopic objects will be subject
to radiative torques unless they are highly symmetric; thus irregular
targets should generally be subject to radiative torques.
Interstellar grains are, of course, comparable to or smaller than
the wavelength of the illuminating radiation, but one does not
expect the radiative torques to vanish when target geometries such as 
Fig. \ref{fig:targs} are 
reduced to sizes comparable to the wavelength.

In this paper we discuss the forces and torques on grains of arbitrary shape
and with sizes which are neither large nor small compared to the wavelength 
of the incident radiation.
We show how 
these forces and torques can be calculated using the discrete dipole
approximation.
Quantitative results are obtained for one particular irregular grain shape.

The first part of this paper,
\S\S\ref{sec:DDA}-\ref{sec:target_orient}, 
is devoted to development of the theory
of scattering by irregular particles to enable efficient 
computation of radiative
torques.
This is carried out
within the conceptual (and computational) framework of
the discrete dipole approximation.
In \S\ref{sec:dda_results} we report results of extensive computations for one
specific irregular grain geometry and composition,
and three different sizes, $a_\eff=0.2$, 0.05, and $0.02\micron$ 
($a_\eff$ is the radius of a sphere of equal volume).

Readers primarily interested in the implications for grain rotation may
elect to skip \S\S\ref{sec:DDA}-\ref{sec:dda_results}, and proceed
directly to \S\ref{sec:interstellar}, where we discuss
the effects of starlight and gas drag under realistic 
conditions in interstellar diffuse clouds,
and \S\ref{sec:superthermal}, where the resulting superthermal rotation is
evaluated.
It is seen in \S\ref{sec:diffuse} 
that moderately anisotropic starlight can torque $a_\eff=0.2\micron$ 
grains in diffuse clouds up to
extremely large rotational velocities,
exceeding even the superthermal rotation due to $\HH$ formation.
Smaller ($a_\eff\ltsim 0.05\micron$) grains, on the other hand, are only
weakly affected by radiative torques.

We also examine the importance of radiative torques within 
quiescent
dark clouds (\S\ref{sec:dark}), and within active star-forming regions
(\S\ref{sec:starforming}).

Our results are summarized in \S\ref{sec:summary}.
Our principal result is that the torques exerted on interstellar grains
by background starlight are dynamically very important for grains
with $a_\eff\gtsim0.1\micron$.
These torques will produce extreme superthermal
rotation of
interstellar grains in both diffuse clouds and star-forming clouds.
In a separate paper (Draine \& Weingartner 1996) we examine the role
of these radiative torques in the alignment of interstellar grains with the
galactic magnetic field.  

\section{The Discrete Dipole Approximation\label{sec:DDA}}

We adopt the ``discrete dipole approximation''
(Draine \& Flatau 1994),
and represent the target by an array of polarizable points,
at locations $\br_j$, with electric polarizabilities $\alpha_j$.
The space between the points is vacuum.
We consider the response of this array to monochromatic illumination,
and adopt the usual complex representation for all time-dependent
variables which are first-order (i.e., $\bE$ and $\bB$ fields, and
dipole moments $\bp_j$).
Second-order variables (i.e., forces $\bF$ and torques $\bGam$) will be taken to
be purely real.
We assume an incident plane wave
\beq
\bE_{inc} = \bE_{inc,0} \exp(i \bk \!\cdot\! \br - i \omega_0 t)
\eeq
\beq
\bB_{inc}=\bB_{inc,0} \exp(i \bk \!\cdot\! \br-i\omega_0 t)
\eeq
\beq
\bB_{inc,0}=\hat{\bk}\times\bE_{inc,0}
\eeq
where the unit vector $\hat{\bk}\equiv\bk/k$, and $k=\omega_0/c$.
By suitable choice of complex $\bE_{inc,0}$ we can represent general
elliptical polarization.

The dipole at location $\br_j$ acquires a dipole moment
\beq
\bp_j(t)=\bp_j(0)e^{-i\omega_0 t}=\alpha_j \bE_j
\eeq
where
$\bE_j$ is the electric field at $\br_j$ due to
the incident electromagnetic field and all dipoles except dipole $j$:
\beq
\bE_j = \bE_{inc,j} + \bE_{sca,j}
\eeq
\beq
\bE_{inc,j}=\bE_{inc,0}\exp(i\bk\!\cdot\!\br_j - i\omega_0 t)
\eeq
\beq
\bE_{sca,j}=
e^{-i\omega_0 t}
\sum_{l\neq j}{e^{ikr_{jl}}\over r_{jl}^3}
\left\{ k^2\br_{jl}\times(\bp_l\times\br_{jl})
+ {(1-ikr_{jl})\over r_{jl}^2}
\left[
3\br_{jl}(\br_{jl}\!\cdot\!\bp_l)-r_{jl}^2\bp_l
\right]
\right\}~~~,
\label{eq:esca}
\eeq
where $\br_{jl}\equiv\br_j-\br_l$.
Similarly, the magnetic field at $\br_j$ may be written
(see, e.g., Jackson 1975).
\beq
\bB_j = \bB_{inc,j} + \bB_{sca,j}
\eeq
\beq
\bB_{inc,j}=
\bB_{inc,0}
%BTD 96.04.16
%e^{i\bk\cdot\br_j - i\omega_0 t}
\exp(i\bk\cdot\br_j - i\omega_0 t)
\eeq
\beq
\bB_{sca,j}=e^{-i\omega_0 t}
\sum_{l\neq j} k^2{e^{ikr_{jl}}\over r_{jl}^2}
(\br_{jl}\times\bp_l)\left(1-{1\over ikr_{jl}}\right)~~~,
\label{eq:bsca}
\eeq
\section{Electromagnetic Force on a Grain}

The instantaneous force on point dipole $j$ is
\beq
\bF_j = 
\left[\Re\left(\bp_j\!\cdot\!\nabla_j\right)\Re(\bE_j) + 
{1\over c}\Re\left({d \bp_j\over dt}\right)\times \Re(\bB_j)
\right] ~~~.
\label{eq:force}
\eeq
The first term in (\ref{eq:force}) is due to gradients in the local electric
field; the second is the Lorentz force on the currents associated with
the oscillating dipole moments.

It is convenient to separate the total force 
$\bF_{rad}=\sum_j\bF_j$ into two terms:
\beq
\bF_{rad} = \bF_{inc} + \bF_{sca} ~~~,
\eeq
\beq
\bF_{inc}=
\sum_{j=1}^N\left[
\Re\left(\bp_j\!\cdot\!\nabla_j\right)
\Re(\bE_{inc,j}) + 
{1\over c}\Re\left({d \bp_j\over dt}\right)\times \Re(\bB_{inc,j})
\right] ~~~,
\eeq
\beq
\bF_{sca}=
\sum_{j=1}^N
\left[\Re\left(\bp_j\!\cdot\!\nabla_j\right)\Re(\bE_{sca,j}) + 
{1\over c}\Re\left({d \bp_j\over dt}\right)\times \Re(\bB_{sca,j})
\right] ~~~.
\label{eq:fsca}
\eeq
The time-averaged force is
\beq
\langle \bF_{rad}\rangle = \langle \bF_{inc}\rangle + \langle\bF_{sca}\rangle ~~~.
\eeq
Letting $x^*$ denote the complex conjugate of $x$, it is straightforward
to show that
\begin{eqnarray}
\langle\bF_{inc}\rangle
&=& {1\over2}\Re
\left[
\sum_{j=1}^N i\bk (\bp_{j}^*(0)\!\cdot\! \bE_{inc,0})\exp(i\bk\!\cdot\!\br_j)
\right] ~~~,
\label{eq:finc}
\\
&=& {1\over 8\pi}C_{ext}|\bE_{inc,0}|^2 \khat ~~~,
\end{eqnarray}
where $C_{ext}$ is the total extinction cross section, evaluated
using the optical theorem (Draine 1988).
Thus $\langle\bF_{inc}\rangle$ is just the 
average rate at which momentum is removed from the
incident radiation field.

Direct evaluation of $\langle \bF_{sca}\rangle$ from
(\ref{eq:fsca}), using (\ref{eq:esca}) and (\ref{eq:bsca}),
involves summing over $N(N-1)$ terms, which for 
$N\simgt10^3$ becomes computationally prohibitive.\footnote{
	These sums could presumably be evaluated in $O(N\ln N)$ operations
	using FFT techniques similar to those used for evaluation of the 
	fields
	$\bE_j$ (Goodman, Draine, \& Flatau 1991), but this has not
	been implemented here.
	}
Alternatively, we 
may evaluate the net rate of momentum transport to infinity by
the scattered radiation field; let this be denoted
$\langle\bF_{out}\rangle$.
Overall momentum conservation requires that the momentum removed from the
incident beam either be transferred to the grain or carried off by the
outgoing scattered radiation:
\beq
{1\over 8\pi}C_{ext}|\bE_{inc,0}|^2\khat =
\langle \bF_{rad} \rangle +
\langle \bF_{out}\rangle ~~~.
\eeq
Thus we see that 
$\langle\bF_{out}\rangle=-\langle\bF_{sca}\rangle$: $\bF_{sca}$ can be
thought of as simply the ``recoil'' of the grain due to the
scattered radiation.

The time-averaged rate of transport of momentum by the scattered radiation
is
\begin{eqnarray}
\langle\bF_{out}\rangle
&=& 
\int r^2 d\Omega {1\over 8\pi} 
\Re\left(\bE_{rad}^*\times\bB_{rad}\right)
\\
&=&
{k^4\over 8\pi}
\int d\Omega ~ \nhat 
\left|
\sum_{j=1}^N \left[\bp_j(0)-\nhat(\nhat\cdot\bp_j(0))\right]
e^{-ik\nhat\cdot\br_j}
\right|^2 ~~~,
\label{eq:fout}
\end{eqnarray}
where in (\ref{eq:fout}) we have assumed the surface of integration
to have radius $r\gg\lambda\equiv 2\pi/k$, retaining only the leading terms
of $O(r^{-1})$
in $\bE$ and $\bB$ (see Appendix A).
Each scattering direction requires evaluation of $O(N)$ terms;
the angular integrations typically require a few hundred scattering directions 
for accurate evaluation.

We define a dimensionless radiation pressure efficiency vector
$\bQ_{pr}$
such that the radiation pressure force
\beq
\langle \bF_{rad}\rangle
=
\bQ_{pr}\pi a_{\rm eff}^2{|\bE_{inc,0}|^2 \over 8\pi} ~~~,
\eeq
where the effective target radius $a_{\rm eff}\equiv(3V/4\pi)^{1/3}$, 
where $V$ is the volume of target material.
The vector $\bQ_{pr}$ depends on both the orientation of the grain
and the direction of propagation and polarization of the incident radiation 
(e.g., linearly or circularly polarized).
It is clear that $\bQ_{pr}\!\cdot\!\khat$ is just the quantity usually 
described as
the ``radiation pressure efficiency factor'' 
$=Q_{abs}+(1-\langle\cos\theta\rangle)Q_{sca}$,
where $Q_{abs}$ and $Q_{sca}$ are the usual absorption and scattering
efficiency factors, and $\langle\cos(\theta)\rangle$ is the mean value
of $\cos\theta$ for the scattered radiation, where $\theta$ is the
scattering angle (see, e.g., Bohren \& Huffman 1983).
It is obvious that 
$\bQ_{pr}\parallel\khat$ for
targets -- e.g., spheres -- which are symmetric 
under rotation around an axis parallel to $\khat$.
\section{Electromagnetic Torque on a Grain\label{sec:torque}}

We choose a coordinate system such that the grain center-of-mass is at
$\br=0$.
The instantaneous radiative torque on the grain is
\beq
\bGam_{rad} =
\sum_{j=1}^N \br_j \times \bF_j + 
\sum_{j=1}^N \Re(\bp_j)\times\Re(\bE_j) ~~~.
\eeq
As for the force calculation, we separate the radiative torque 
into ``incident'' and ``scattered'' components:
\beq
\bGam_{rad} = 
\bGam_{inc} + \bGam_{sca}~~~;
\eeq
$\bGam_{inc}$ is the torque exerted on the oscillating dipoles by 
$\bE_{inc}$ and $\bB_{inc}$, and
$\bGam_{sca}$ is the torque exerted on the oscillating dipoles by the
scattered radiation field -- i.e., the fields 
$\bE_{sca}$ and $\bB_{sca}$ due to the oscillating
dipoles.
The time-averaged value of $\bGam_{inc}$ is straightforward:
\beq
\langle\bGam_{inc}\rangle=
{1\over2}\Re\left[
\sum_{j=1}^N
\bp_j^*(0)\times\bE_{inc,0}e^{i\bk\!\cdot\!\br_j}
-i\bk\times\sum_{j=1}^N
\br_j\left(\bp_j^*(0)\!\cdot\!\bE_{inc,0}\right)e^{i\bk\cdot\br_j}
\right]~~~.
\eeq
As was the case for $\bF_{sca}$,
direct evaluation of $\bGam_{sca}$ would be time-consuming.
By analogy to $\bF_{sca}$, 
we appeal to the fact that the time-averaged torque exerted by the
scattered field plus the time-averaged
rate of transport of angular momentum to infinity
by this field
must sum to zero:
$\langle\bGam_{sca}\rangle+\langle\bGam_{out}\rangle=0$, where
$\bGam_{out}$ is the net rate of transport of angular momentum
by the scattered radiation field across a fixed surface enclosing the
target.
In dyadic notation, the instantaneous angular momentum flux may be
written (Jackson 1975, problem 6.11)
\beq
\bM={1\over 4\pi}
\left(
\Re(\bE)\Re(\bE) + 
\Re(\bB)\Re(\bB) -
{|\Re(\bE)|^2+|\Re(\bB)|^2\over2}\bI
\right)\times\br ~~~,
\eeq
where $\bI$ is the unit dyadic.
Integrating $\bM$ over a spherical surface of
radius $r$:
\beq
\bGam_{out}=\int_S r^3{d\Omega\over 4\pi}
\nhat\!\cdot\!
\left[
\Re({\bE})\Re({\bE}) + \Re(\bB)\Re(\bB) 
\right]\times\nhat ~~~.
\label{eq:gamout}
\eeq
Now, if $r\gg\lambda$, the field radiated by the oscillating
dipoles has 
$|\nhat\!\cdot\!\bE_{sca}|\propto r^{-2}$,
$|\nhat\!\cdot\!\bB_{sca}|\propto r^{-2}$,
$|\bE_{sca}\times\nhat|\propto r^{-1}$, and
$|\bB_{sca}\times\nhat|\propto r^{-1}$.
Thus we need retain only the leading-order terms in evaluating each of
these quantities.
Expanding $\bE_{sca}$ and $\bB_{sca}$ as shown in Appendix A, 
letting $r\rightarrow\infty$,
\beq
\langle\bGam_{sca}\rangle =
-\langle\bGam_{out}\rangle
=
-{k^4\over 8\pi}
\int d\Omega ~\Re\left(S_E^*\bV_B + S_B^*\bV_E\right) ~~~,
\label{eq:gamout2}
\eeq
\beq
S_E \equiv \sum_{j=1}^N
\left[ \br_j - (\nhat\!\cdot\!\br_j)\nhat - {2i\over k}\nhat
\right]\!\cdot\!\bp_j(0) \exp(-ik\nhat\!\cdot\!\br_j) ~~~,
\eeq
\beq
S_B \equiv \nhat\!\cdot\!\sum_{j=1}^N
\bp_j(0)\times \br_j \exp(-ik\nhat\!\cdot\!\br_j) ~~~,
\eeq
\beq
\bV_E \equiv \sum_{j=1}^N 
\left[ \bp_j(0) - \nhat(\nhat\!\cdot\!\bp_j(0)) \right] 
\exp(-ik\nhat\!\cdot\!\br_j) ~~~, 
\eeq
\begin{eqnarray}
\bV_B &\equiv& \sum_{j=1}^N \bp_j(0)\times\nhat\exp(-ik\nhat\!\cdot\!\br_j) ~~~,
\\
&=&-\nhat\times\bV_E ~~~.
\end{eqnarray}
Noting that for pure right circularly polarized radiation, the time-averaged
angular momentum flux is just 
$|\bE_{inc,0}|^2\khat/8\pi k$ (easily seen by noting that each
photon carries angular momentum $\hbar$),
it is convenient to define a dimensionless ``torque efficiency vector''
$\bQ_\Gamma\equiv\bQ_{\Gamma,inc}+\bQ_{\Gamma,sca}$
such that the time-averaged torque
\beq
\langle\bGam_{rad}\rangle = \langle\bGam_{inc}\rangle + \langle\bGam_{sca}\rangle
=
\bQ_\Gamma \pi a_\eff^2 
{|\bE_{inc,0}|^2 \over 8\pi k}
=\pi a_\eff^2 u_{rad} {\lambda \over 2\pi} \bQ_\Gamma ~~~,
\eeq
where $u_{rad}=|\bE_{inc,0}|^2/8\pi$ 
is the time-averaged energy density of the incident
radiation.  Clearly we have
\beq
\bQ_{\Gamma,inc} = {4k\over a_\eff^2 |\bE_{inc,0}|^2}
\Re\left[
\sum_{j=1}^N
\bp_j^*(0)\times\bE_{inc,0}e^{i\bk\cdot\br_j}
-i\bk\times\sum_{j=1}^N
\br_j\left(\bp_j^*(0)\!\cdot\!\bE_{inc,0}\right)e^{i\bk\cdot\br_j}
\right]~~~,
\eeq
\beq
\bQ_{\Gamma,sca} = {-k^5 \over \pi a_\eff^2 |\bE_{inc,0}|^2}
\int d\Omega ~\Re(S_E^* \bV_B + S_B^* \bV_E) ~~~.
\eeq
The direction and magnitude of $\bQ_\Gamma$ depends on the grain
orientation, and on the direction of propagation and
polarization
of the incident radiation.

It is clear that for right circularly polarized radiation 
with $\lambda \gg a_{\rm eff}$, we will have 
$\bQ_{\bGam}\approx Q_{abs}\khat$.

\section{Target Orientation\label{sec:target_orient}}

It is necessary to specify the orientation of the target relative to
the incident radiation.
Let the target contain two axes $\ahat_1$ and $\ahat_2$ (with
$\ahat_1\!\cdot\!\ahat_2=0$) which are fixed relative to the target.
Let us further suppose that $\ahat_1$,
$\ahat_2$, and $\ahat_3=\ahat_1\times\ahat_2$
are the principal axes of the target's moment of inertia tensor, with
moments of inertia $I_1\geq I_2\geq I_3$ about these axes.
The grain orientation is completely specified by the
orientations of any two non-parallel fixed axes, 
in particular $\ahat_1$ and $\ahat_2$.

\begin{figure}
\epsscale{1.00}
\plotone{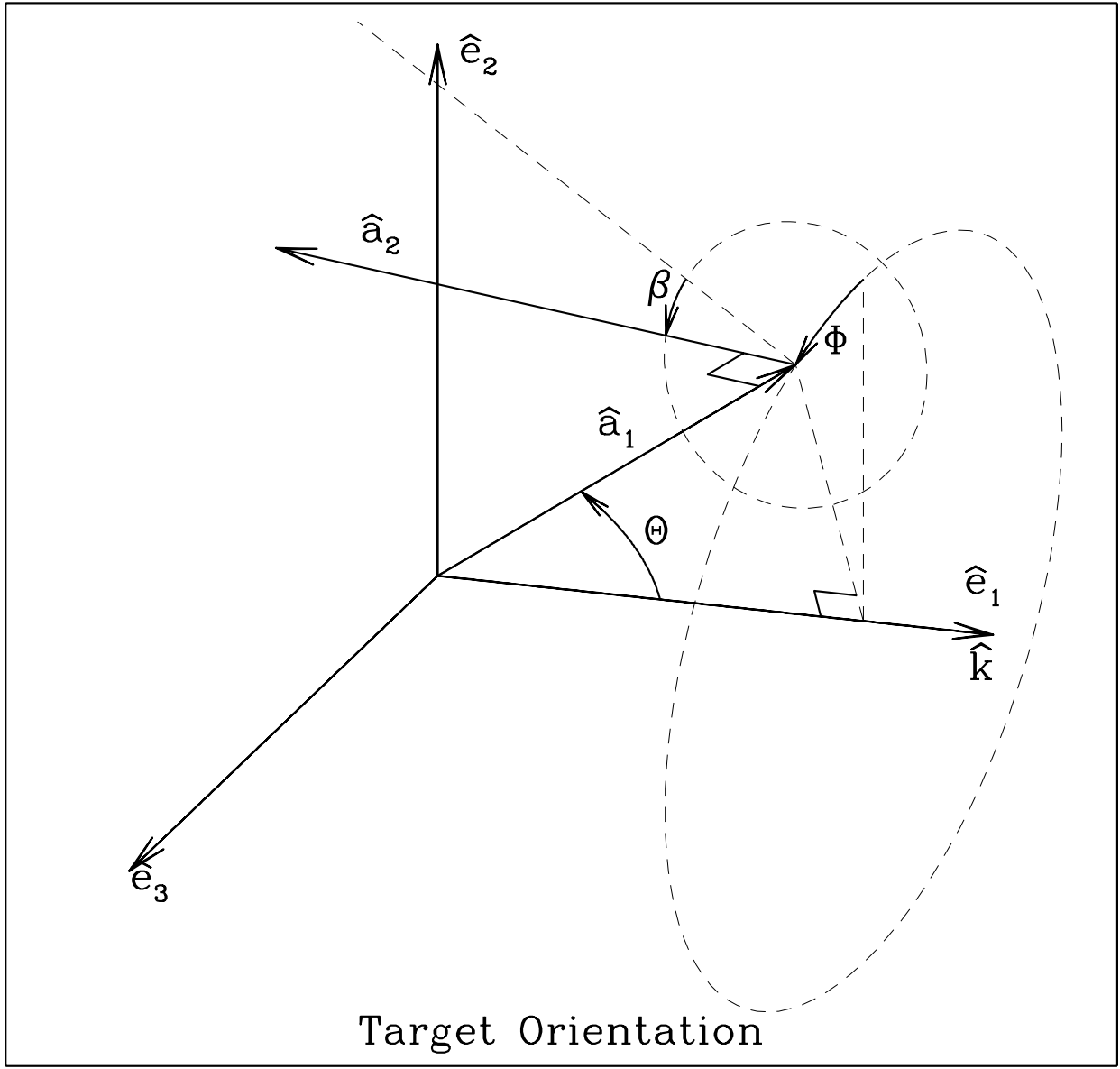}
\caption{
	Grain orientation in ``scattering coordinates''
	\label{fig:orient}
	}
\end{figure}

We define a ``scattering frame'', defined by unit vectors
$\hat{\be}_1=\hat{\bk}$, $\hat{\be}_2\perp\hat{\be}_1$
and $\hat{\be}_3=\hat{\be}_1\times\hat{\be}_2$.
Three angles are required to specify the target orientation.
The orientation of $\ahat_1$ in the scattering frame is described by 
the two angles
$\Theta\in[0,\pi]$ and $\Phi\in[0,2\pi]$,
where
\beq
\ahat_1 = \cos\Theta \ehat_1 + \sin\Theta\cos\Phi\ehat_2 + 
\sin\Theta\sin\Phi\ehat_3
\eeq
A third angle $\beta\in[0,2\pi]$ describes rotations of the target 
axis $\ahat_2$ around $\ahat_1$:
\beq
\ahat_2 = -\sin\Theta\cos\beta\hat{\be}_1
+ (\cos\Theta\cos\beta\cos\Phi-\sin\beta\sin\Phi)\hat{\be}_2
+ (\cos\Theta\cos\beta\sin\Phi+\sin\beta\cos\Phi)\hat{\be}_3
\eeq
The target orientation in the scattering frame is illustrated in 
Fig.\ \ref{fig:orient}.

\section{Computations for One Shape\label{sec:dda_results}}

\begin{figure}
\epsscale{1.00}
\plotone{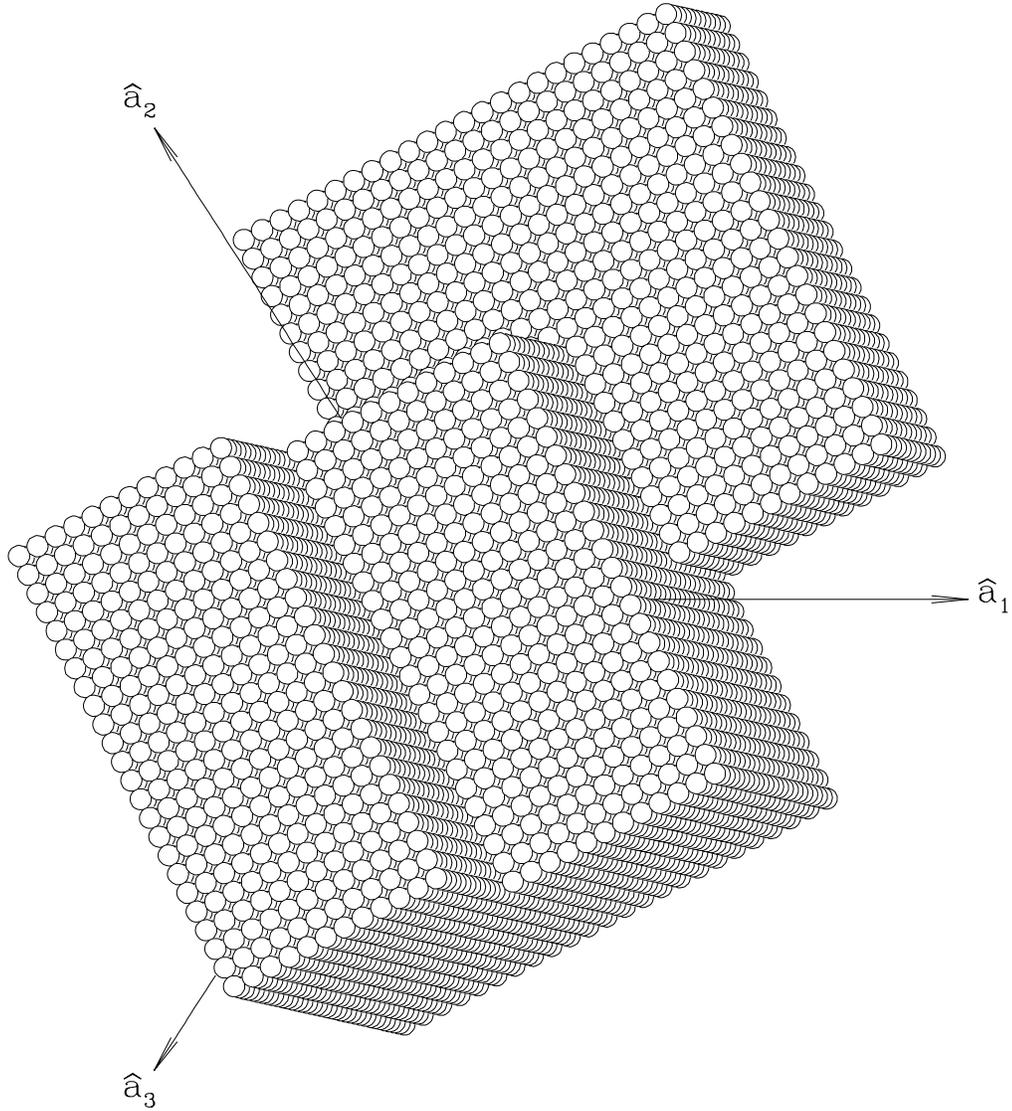}
\caption{
	Representative irregular grain shape.
	The principal axes $\ahat_1$, $\ahat_2$, and $\ahat_3$
	are shown.
	\label{fig:shape}
	}
\end{figure}

We consider one specific grain shape.
This shape, shown in Fig.\ \ref{fig:shape}, is not intended to be
realistic, but is chosen to be (1) easily described, (2) asymmetric,
and (3) well-suited to representation by a cubic array of point dipoles.
The shape chosen is an assembly of thirteen identical cubes.
The coordinates of these cubes (in units of cube width) are listed 
in Table \ref{tab:cubes}.
The first eight cubes are arranged to form a single larger cube;
the remaining five
cubes are attached to the central eight cubes.

Let the eigenvalues $I_1\geq I_2\geq I_3$ of the moment of inertia tensor
be written
\beq
I_j = {2\over5}\alpha_j M a_\eff^2 = {8\pi\over 15}\alpha_j \rho a_\eff^5 ~~~;
\eeq
where $\rho$ is the solid density, and $\alpha_j$ are geometric factors.
A sphere has $\alpha_j=1$;
a 2:2:1 rectangular solid (``brick'') has 
$\alpha_1=(5/3)(\pi/3)^{2/3}=1.719$, 
$\alpha_2=\alpha_3=(25/24)(\pi/3)^{2/3}=1.074$.
The principal axes $\ahat_j$ and factors $\alpha_j$ for our adopted grain
geometry are given in Table \ref{tab:axes}.

Let $T_{gr}$ be the grain temperature.
Purcell (1979) has shown that
if the grain is spinning with angular momentum $\bJ$,
and $J^2\gg I_1 kT_{gr}$, 
then viscoelastic
damping and Barnett effect damping will rapidly bring the grain's
principal axis $\ahat_1$ into alignment with $\bJ$.
In our discussion below we will therefore assume that $\ahat_1$ is at all
times perfectly aligned with the angular momentum $\bJ$ of the
grain.
The grain rotation period will be short compared to all other dynamical
time scales, so it is appropriate to average the forces and torques
over the grain rotation angle $\beta$, and henceforth we will discuss
only the ``$\beta$-averaged'' efficiency vectors 
$\bQ_{pr}(\Theta,\Phi)$ and $\bQ_{\bGam}(\Theta,\Phi)$.

We have carried out scattering calculations for grains with the shape
shown in Fig.\ \ref{fig:shape},
the refractive index of ``astronomical silicate'' 
(Draine \& Lee 1984)\footnote{
	This dielectric function may be obtained by anonymous ftp to
	astro.princeton.edu, file ``draine/dust/eps.Sil''.
	}
and effective radii
$a_\eff = 0.2$, $0.05$, and $0.02\micron$.
The computations were carried out using the discrete dipole approximation
program DDSCAT.5a\footnote{
	The Fortran program DDSCAT.5a is available from 
	B.T.\ Draine and P.J.\ Flatau; contact draine@astro.princeton.edu
	for further information.
	}.
The targets were represented by arrays of $N=6656$, $22464$,
and $53248$ dipoles;
in each case, $N$ was taken to be large enough to satisfy the
validity criterion $|m|kd < 1$ (Draine \& Flatau 1994),
where $m(\omega)$ is the complex refractive index, and
$d=(4\pi/3N)^{1/3}a_\eff$ is the interdipole spacing.
Lattice dispersion relation theory (Draine \& Goodman 1993) was used
to determine the optimal choice of dipole polarizabilities at each
wavelength.
The computations employed FFT techniques (Goodman, Draine \& Flatau 1991)
and a stabilized version of the bi-conjugate-gradients method
(Flatau 1996); nevertheless the computational demands were substantial,
because of the need to compute scattering for many different
wavelengths and orientations.
For example,
determining the scattering properties for one orientation and
two polarization states for the $a_\eff=0.2\micron$ target at
a single wavelength
$\lambda=0.4\micron$, with the target represented by $N=22464$ dipoles,
required 720 cpu-sec on a Sun-10 computer with a
125MHz Hypersparc cpu.
For the $a_\eff=0.2\micron$ grain, 
computations were carried out for 
26 distinct wavelengths (from $0.1\micron$ to $20\micron$) and
285 orientations (15 values of $\beta$ for each of 19 values of $\Theta$),
requiring $\sim2000$ cpu-hr.

\begin{figure}
\epsscale{1.00}
\plotone{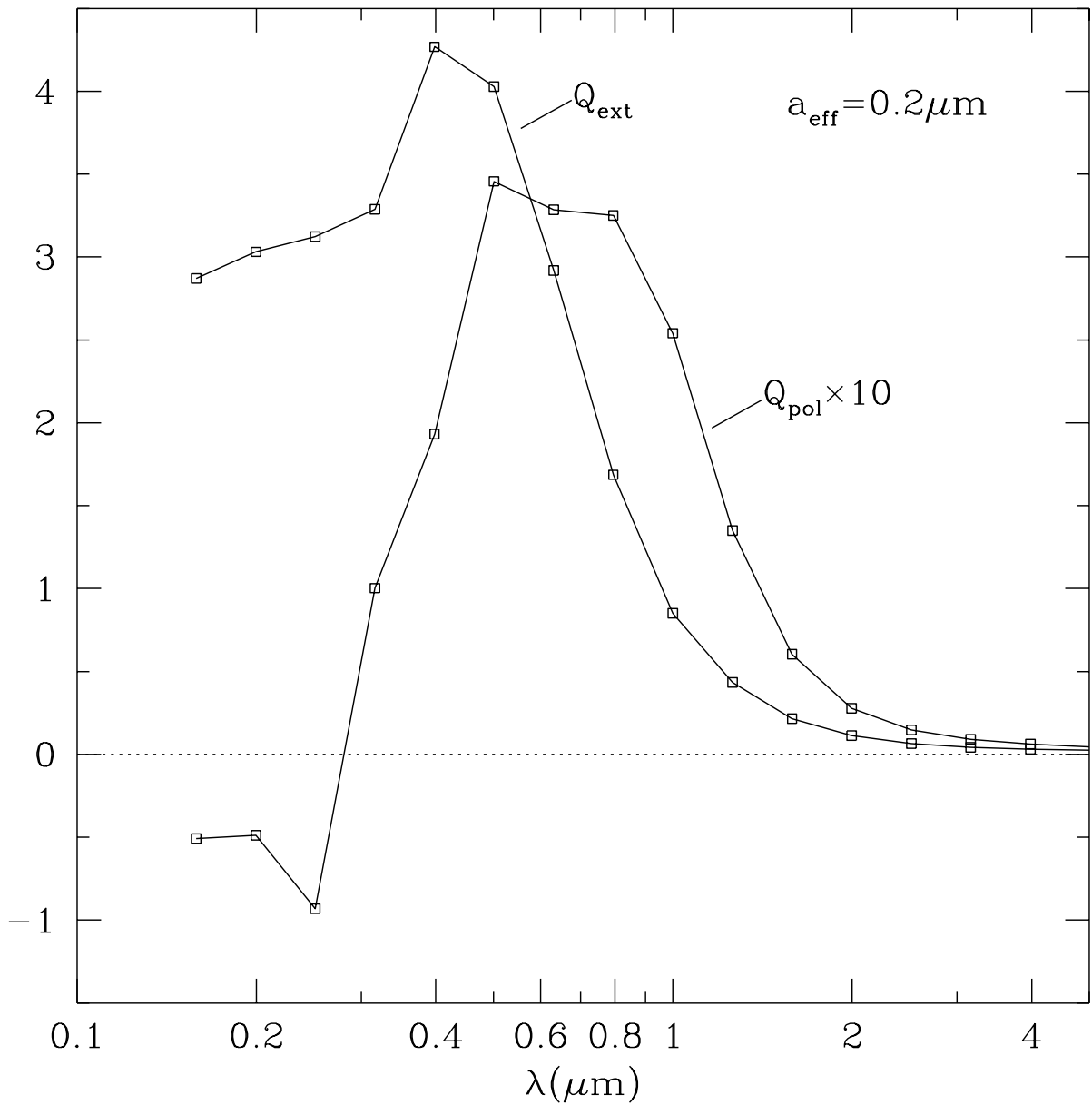}
\caption{
	Extinction and polarization efficiency factors for 
	the grain shape shown in Fig.~\protect{\ref{fig:shape}},
	with $a_\eff=0.2\micron$,
	spinning around the principal axis $\ahat_1$,
	with radiation propagating perpendicular to $\ahat_1$.
	The polarization cross section peaks at $\lambda=0.5\micron$,
	with $Q_{pol}/Q_{ext}=0.086$ for this particular grain.
	\label{fig:qpol_0.2}
	}
\end{figure}

\begin{figure}
\epsscale{1.00}
\plotone{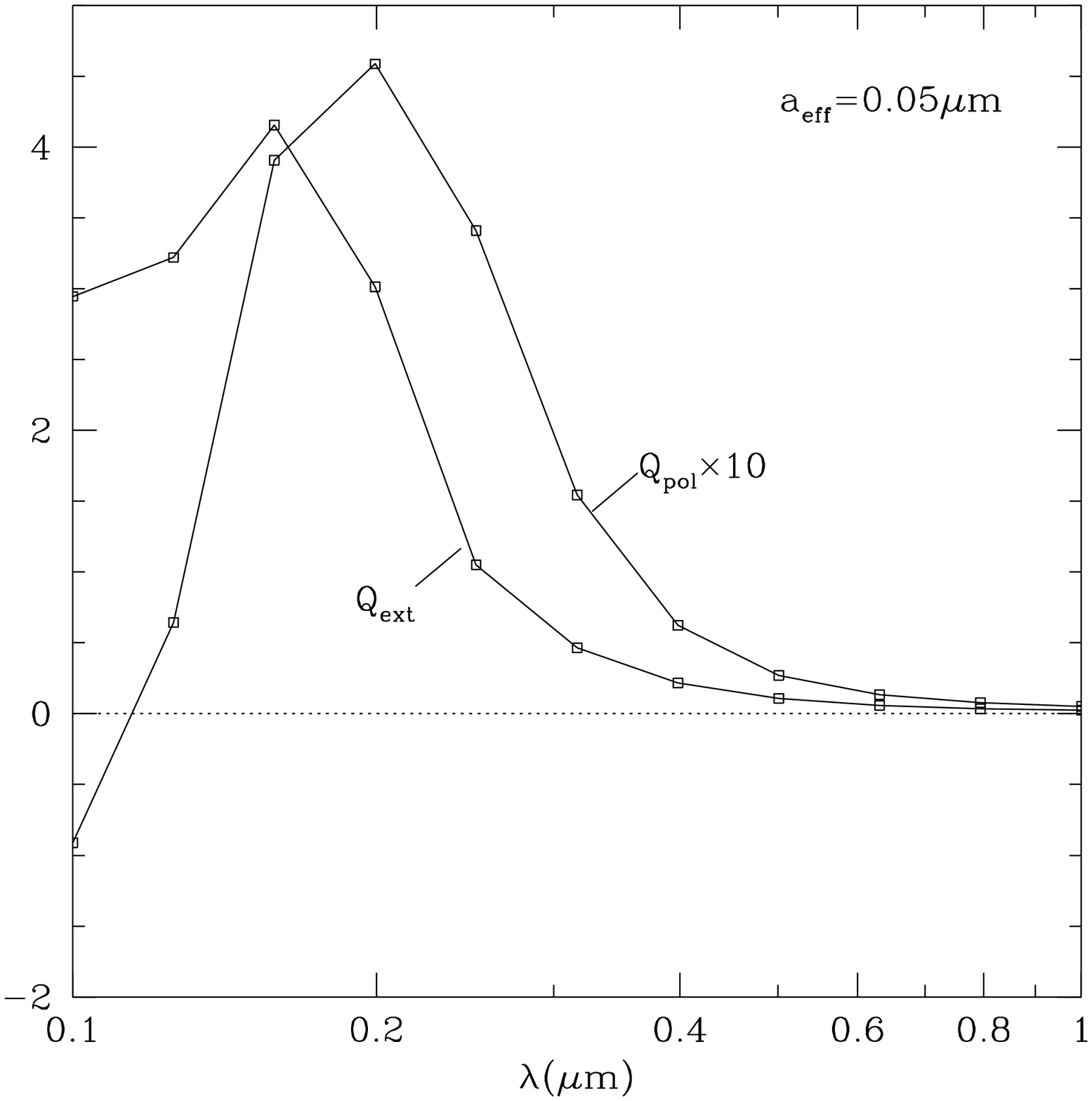}
\caption{
	Same as Fig.~\protect{\ref{fig:qpol_0.2}}, but for
	$a_\eff=0.05\micron$, with $Q_{pol}$ peaking at
	$\lambda=0.2\micron$, with $Q_{pol}/Q_{ext}=0.15$.
	\label{fig:qpol_0.05}
	}
\end{figure}

Consider a grain spinning around axis $\ahat_1$, with $\ahat_1\perp\bk$.
In Fig.\ \ref{fig:qpol_0.2} we show 
\beq
Q_{ext}\equiv {C_{ext}(\bE\perp \ahat_1)+C_{ext}(\bE\parallel\ahat_1)\over
2\pi a_\eff^2}
\eeq
\beq
Q_{pol}\equiv {C_{ext}(\bE\perp \ahat_1)-C_{ext}(\bE\parallel\ahat_1)\over
\pi a_\eff^2}
\eeq
where
$C_{ext}$ is the extinction cross section averaged over grain rotation
angle $\beta$.
We see that the polarization cross
section for the $a_\eff=0.2\micron$ grain peaks near $\lambda=0.5\micron$.
Grains of approximately this size would therefore be suitable for
producing the bulk of the observed interstellar polarization, which
tends to peak near $\lambda\approx0.55\micron$.
The $a_\eff=0.05\micron$ grain, on the other hand, is most effective
at polarizing near $\lambda=0.2\micron$ (see Fig.\ \ref{fig:qpol_0.05});
the observed wavelength-dependence of interstellar polarization
indicates that grains of this size are {\it not} appreciably
aligned in the interstellar medium (Kim \& Martin 1994b).

We can restrict our calculations to $\Phi=0$; for other values of $\Phi$,
the efficiency vectors $\bQ$ may be obtained by a simple
rotation:
\beq
\bQ(\Theta,\Phi)=
\bQ(\Theta,0)\!\cdot\!\ehat_1\ehat_1 +
\bQ(\Theta,0)\!\cdot\!\ehat_2(\ehat_2\cos\Phi+\ehat_3\sin\Phi) +
\bQ(\Theta,0)\!\cdot\!\ehat_3(\ehat_3\cos\Phi-\ehat_2\sin\Phi)~~~.
\eeq

\begin{figure}
\epsscale{1.00}
\plotone{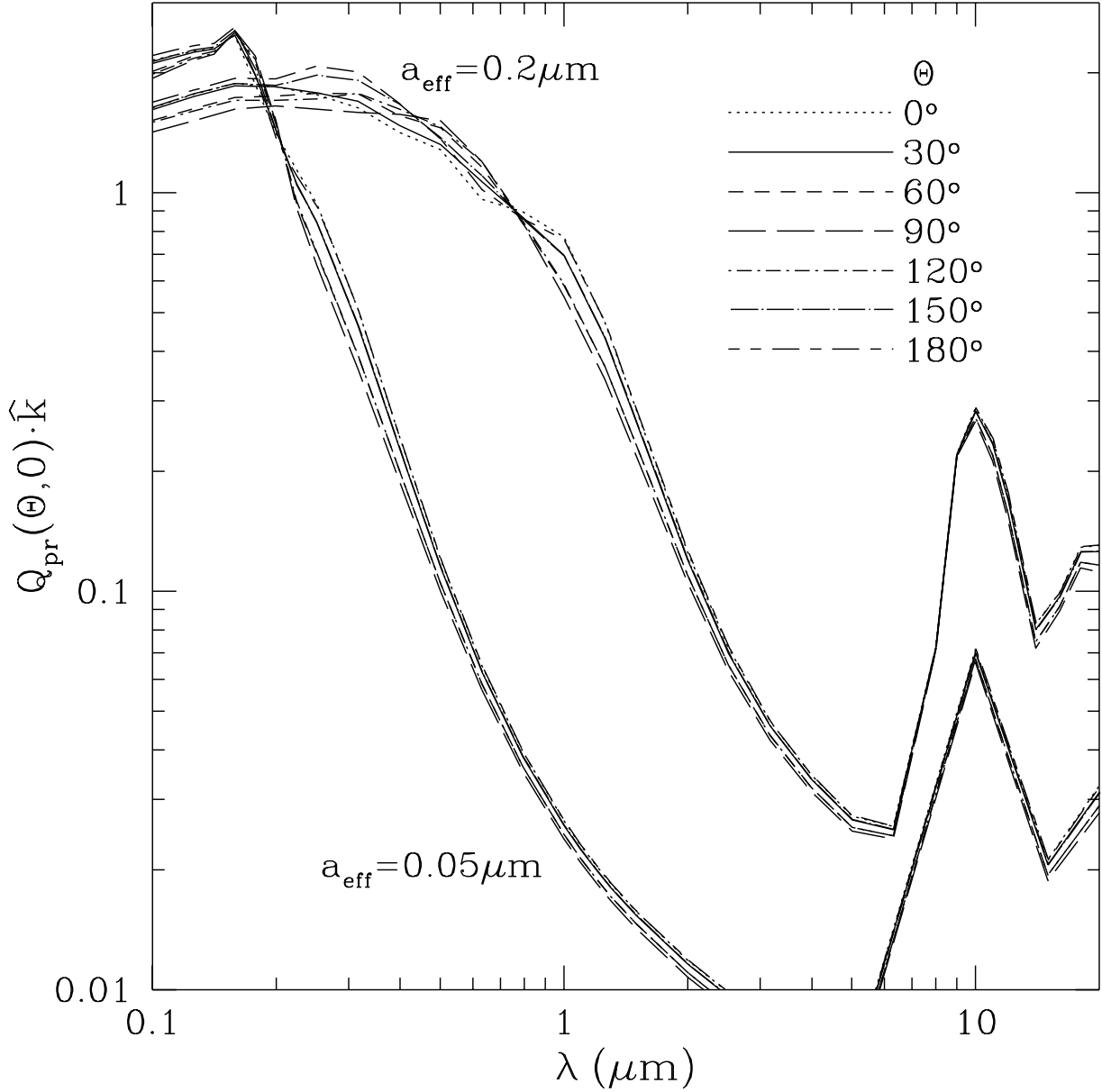}
\caption{
	\label{fig:qpr_kvslambda}
	The component of the radiation pressure efficiency vector 
	$\bQ_{pr}(\Theta,0)$
	along $\khat$, the propagation direction of the incident radiation,
	as a function of wavelength $\lambda$, for several values of the
	angle $\Theta$ between $\khat$ and 
	the grain axis $\ahat_1$, and for two different grain sizes,
	$a_{\rm eff}=0.2\micron$ and $0.05\micron$.
	$\bQ_{pr}$ is averaged over the angle $\beta$ measuring
	rotations of the grain around $\ahat_1$.
	The peak near $10\micron$ is the silicate absorption feature.
	}
\end{figure}

Let ($\bQ_{pr}^1$, $\bQ_{pr}^2$) and ($\bQ_{\bGam}^1$, $\bQ_{\bGam}^2$) 
be the radiation pressure and radiation torque
efficiency vectors for polarization states 1 and 2 (these can be
be any two orthogonal polarization states, e.g., two linear polarization
states, or left- and right-handed circular polarization states).
For unpolarized incident radiation (i.e., equal amounts of two orthogonal
polarization states) we have efficiency vectors
\begin{eqnarray}
\bQ_{pr}^{un}={1\over2}(\bQ_{pr}^1+\bQ_{pr}^2)
\\
\bQ_{\bGam}^{un}={1\over2}(\bQ_{\bGam}^1+\bQ_{\bGam}^2)~~~.
\end{eqnarray}
Henceforth in this paper we will be concerned only with unpolarized
radiation, and we will omit the $^{un}$ superscript.
In Fig.\ \ref{fig:qpr_kvslambda} we show our results for the component of
$\bQ_{pr}(\Theta,0)$ along $\khat$, as a function of $\lambda$,
for several values of $\Theta$.
This is, of course, just the usual ``radiation pressure efficiency
factor'' $[Q_{abs}+(1-\langle\cos\Theta\rangle)Q_{sca}]$.
We see the expected tendency for $\bQ_{pr}\!\cdot\!\khat$ to be small when
$\lambda\gg a_\eff$, and to be of order unity when
$2\pi a_\eff/\lambda \gtsim 1$ (i.e., $\lambda\ltsim 1\micron$ for 
$a_\eff=0.2\micron$, and $\lambda\ltsim 0.3\micron$ for $a_\eff=0.05\micron$.

\begin{figure}
\epsscale{1.00}
\plotone{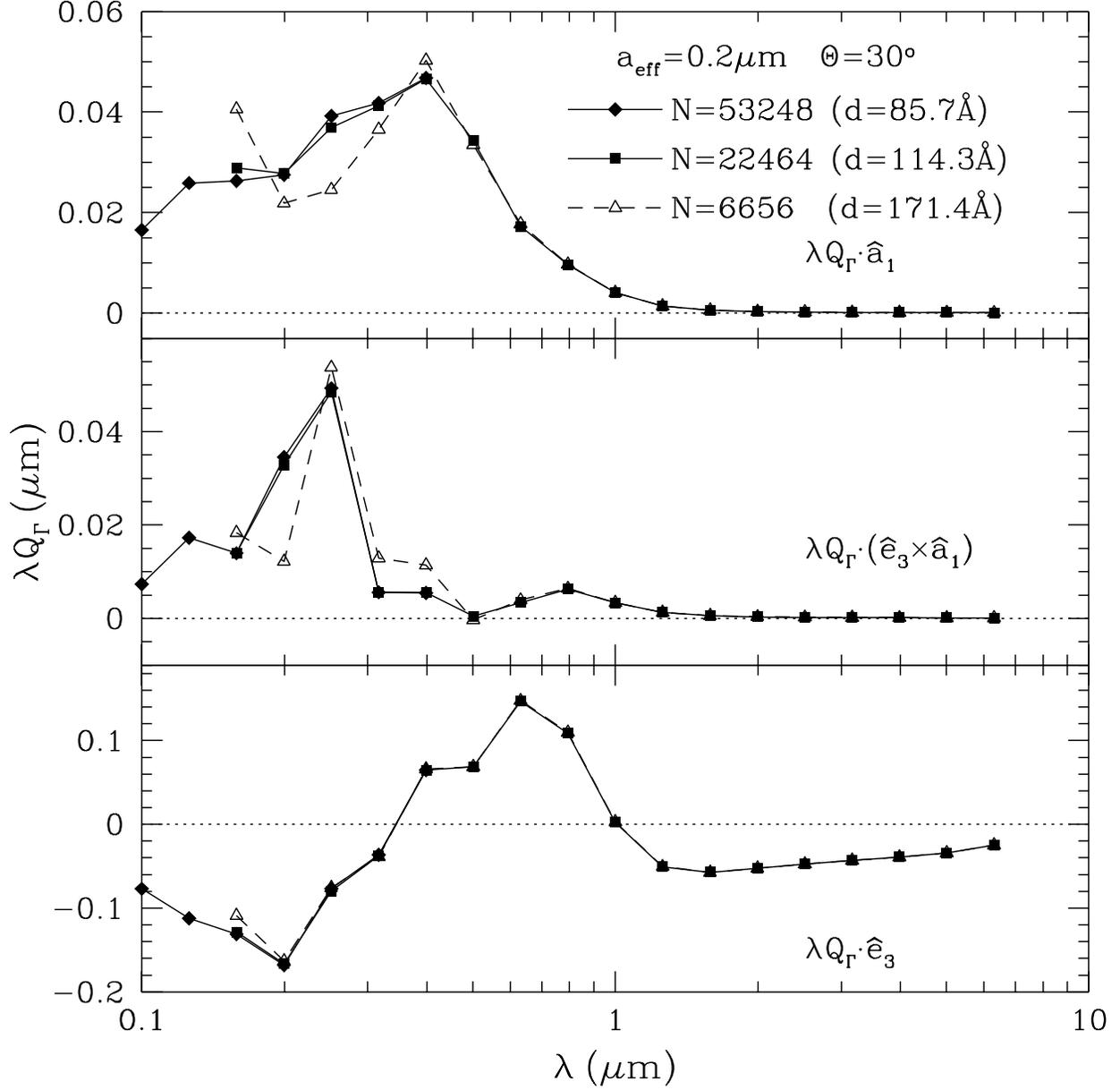}
\caption{
	\label{fig:qgam_0.2_30}
	Three components of $\lambda \bQ_\Gamma$ for $\Theta=30^o$,
	as a function of $\lambda$, for $a_\eff=0.2\micron$ grain.
	Results are shown for DDA calculations using
	$N=6656$, $22464$, and $53248$ dipoles.
	It is seen that the $N=22464$ dipole array provides a
	good approximation to the target for $\lambda\gtsim0.15\micron$.
	}
\end{figure}

In Fig.\ \ref{fig:qgam_0.2_30} we show the three components of the radiation
torque efficiency vector $\bQ_\Gamma$ 
for the $a_\eff=0.2\micron$ target oriented
at $\Theta=30\arcdeg$.
For $\lambda\geq0.1585\micron$ results are shown for three different
dipole arrays, with $N=6656$, 22464, and 53248 dipoles, with interdipole
separations $d=171$, 114, and $85.7\Angstrom$.
At
%BTD 96.04.16 this wavelength 
$\lambda=0.1585\micron$ 
we have $m=2.153+0.263i$, so that for the three dipole
arrays at this wavelength we have $|m|kd=1.47$, 0.98, and 0.74.
We see that the DDA calculations 
%BTD 96.04.16
appear to 
have converged when $|m|kd<1$.
At the shortest wavelength $\lambda=0.1\micron$ we have
$m=1.545+0.956i$, and for the $N=53248$ array we have $|m|kd=0.98$, marginally
satisfying our validity criterion.

\begin{figure}
\epsscale{1.00}
\plotone{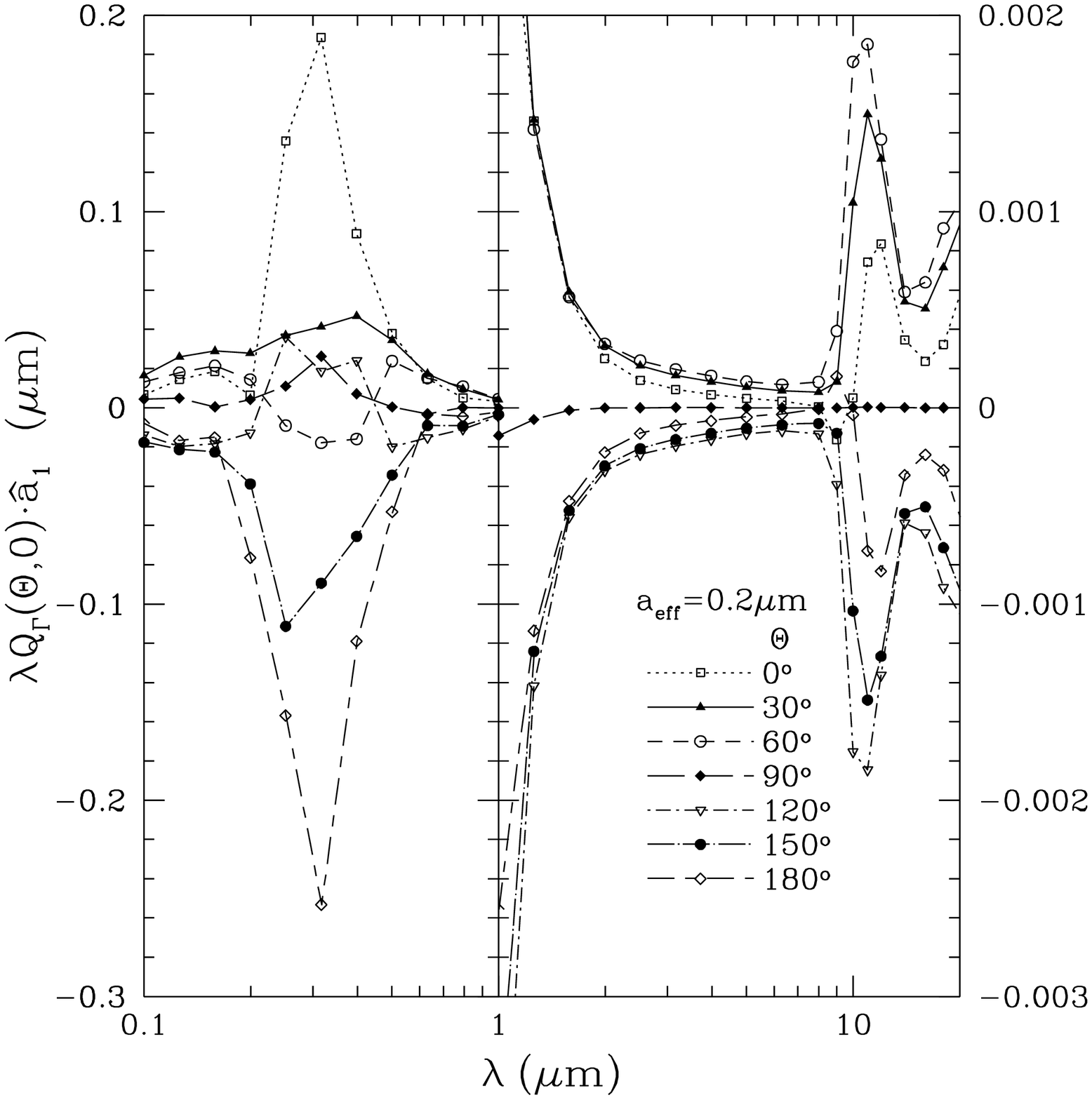}
\caption{
	\label{fig:qgam_a1vslambda_0.2}
	$\lambda\bQ_{\bGam}(\Theta,0)\!\cdot\!\ahat_1$,
	where $\bQ_{\bGam}$ is the radiation torque efficiency 
	vector and $\ahat_1$ is the principal axis with the
	largest moment of inertia,
	for $a_{\rm eff}=0.2\micron$ grain and various values
	of the angle $\Theta$ between $\ahat_1$ and $\khat$.
	}
\end{figure}

\begin{figure}
\epsscale{1.00}
\plotone{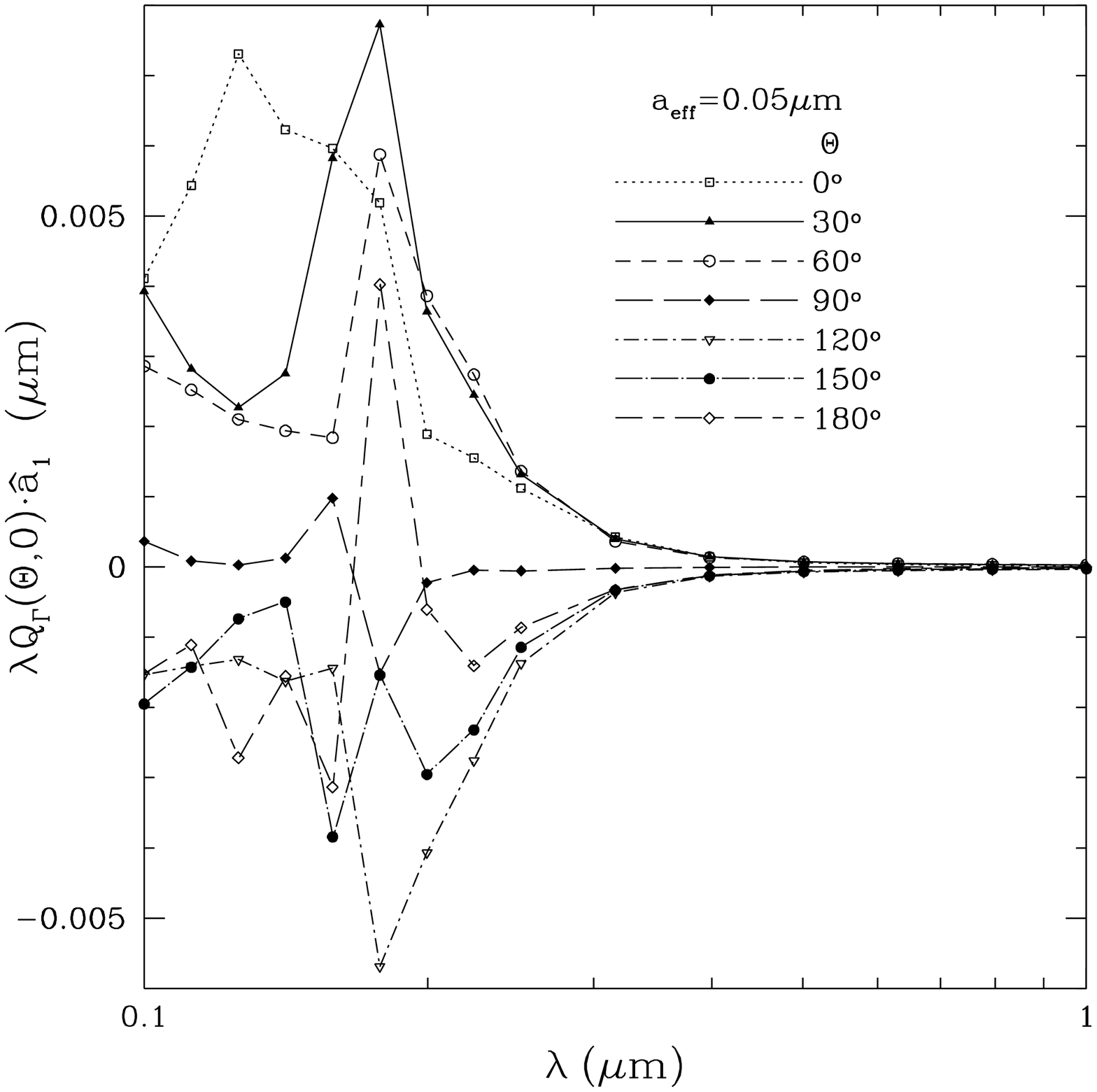}
\caption{
	Same as Fig. \protect{\ref{fig:qgam_a1vslambda_0.2}}
	but for $a_{\rm eff}=0.05\micron$.
	\label{fig:qgam_a1vslambda_0.05}
	}
\end{figure}

Figs.\ \ref{fig:qgam_a1vslambda_0.2} and 
\ref{fig:qgam_a1vslambda_0.05} show, as a function of $\lambda$,
the component of $\bQ_\Gamma$ along the principal axis $\ahat_1$ about which
we have assumed the grains to be spinning, for
$a_\eff=0.2\micron$ and $0.05\micron$ grains.
This component of $\bQ_\Gamma$ will tend to spin up 
the grain if $\bQ_\Gamma\!\cdot\!\ahat_1/\bJ\!\cdot\!\ahat_1 >0$, or to spin it
down if $\bQ_\Gamma\!\cdot\!\ahat_1/\bJ\!\cdot\!\ahat_1 <0$.

\section{Interstellar Grains\label{sec:interstellar}}

\subsection{Radiation Fields}

The average interstellar radiation field (``ISRF'') spectrum in
the solar neighborhood has been estimated to be
(Mezger, Mathis, \& Panagia 1982; Mathis, Mezger, \& Panagia 1983)
\begin{eqnarray}
\lambda u_{\ISRF,\lambda}
&=&
U(\lambda) ~~~~~{\rm for~}\lambda < 2460\Angstrom
\\
&=&
{4\pi\lambda\over c}
\sum_{i=1}^3 W_i B_\lambda(T_i) ~~~~~{\rm for~}\lambda>2460\Angstrom
\end{eqnarray}
where the ultraviolet function
\begin{eqnarray}
U(\lambda)
&=&0~~~~~~~~~~~~~~~~~~~~~~~~~~~~~~~~~{\rm ~for~}\lambda < 912\Angstrom
{\rm ~or~}\lambda > 2460\Angstrom
\nonumber
\\
&=&1.287\times 10^{-9}(\lambda/\micron)^{4.4172} 
\erg\cm^{-3}~~~~~{\rm ~for~}912-1100\Angstrom
\nonumber
\\
&=&6.825\times 10^{-13}(\lambda/\micron) 
\erg\cm^{-3}~~~~~~~~~~{\rm ~for~}1100-1340\Angstrom
\nonumber
\\
&=&2.373\times 10^{-14}(\lambda/\micron)^{-0.6678} 
\erg\cm^{-3}~~~~~{\rm ~for~}1340-2460\Angstrom ~~~.
\label{eq:uvfield}
\end{eqnarray}
The three blackbody components are indicated in Table \ref{tab:radfield},
where
we also provide values of $u$ and $\lambdabar$ for each of the four
spectral components, where
\beq
u\equiv\int u_\lambda d\lambda ~~~,
\eeq
\beq
\lambdabar \equiv
{\int \lambda u_\lambda d\lambda
\over
\int u_\lambda d\lambda} ~~~.
\eeq
The total starlight energy density is
$u_\ISRF=8.64\times10^{-13}\erg\cm^{-3}$.
We expect appreciable anisotropy in the typical interstellar radiation field:
generally, there will be more starlight from the direction of the
Galactic center, the nearest bright star or association may 
be important, and nearby dust may attenuate the starlight over one
part of the sky.
We define an anisotropy parameter $\gamma$, such that
the net Poynting flux 
is $\gamma u_{rad}c$,
where $u_{rad}$ is the total energy density of radiation.
We 
%BTD 96.04.16
%estimate
assume
$\gamma\approx 0.1$ to be a characteristic value.
In our discussions below we will represent the radiation field
as an isotropic component with energy density $(1-\gamma)u_{rad}$
plus 
a unidirectional component with energy density $\gamma u_{rad}$.

\subsection{Radiative Forces and Torques due to Starlight}

\subsubsection{Unidirectional Radiation}
For a grain subject to a unidirectional radiation field with energy
density $\gamma u_{rad}$, the force and
torque on the grain are
\begin{eqnarray}
\bF_{rad} &=& \pi a_\eff^2 \gamma u_{rad} \langle \bQ_{pr}(\Theta,\Phi)\rangle
\\
\bGam_{rad} &=& \pi a_\eff^2 ~\gamma u_{rad}~ {\lambdabar\over2\pi}~
\langle \bQ_\Gamma(\Theta,\Phi)\rangle
\end{eqnarray}
where 
$\langle\rangle$ now denotes spectral averaging:
\beq
\langle \bQ(\Theta,\Phi) \rangle \equiv 
{\int \bQ(\Theta,\Phi)\lambda u_\lambda d\lambda
\over
\int \lambda u_\lambda d\lambda} ~~~.
\eeq

In Table \ref{tab:grain_props}
we present values of $\ahat_1\!\cdot\!\langle\bQ_{pr}\rangle$ and
$\ahat_1\!\cdot\!\langle\bQ_{pr}\rangle$ for
$(\Theta,\Phi)=(0,0)$, with the spectral averages evaluated for
each of the $u_\lambda$ in Table \ref{tab:radfield},
for three grain sizes, $a_\eff=0.02$, 0.05, and $0.2\micron$.
Note that the torque efficiencies $\langle\bQ_\Gamma\rangle$
become very small as the grain size is decreased below
$\sim0.1\micron$.

\begin{figure}
\epsscale{1.00}
\plotone{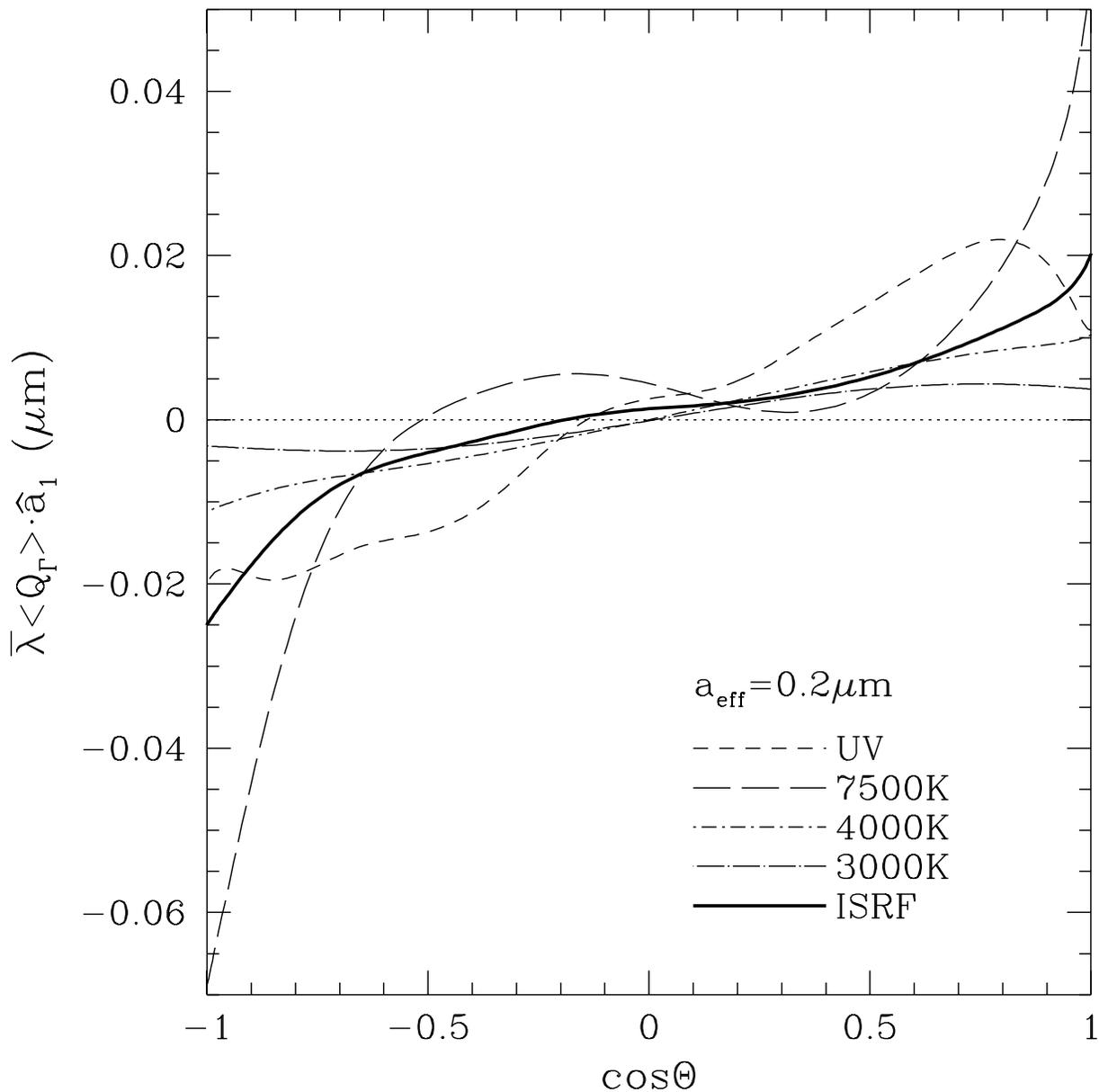}
\caption{
	\label{fig:qgam_a1vsTheta_0.2}
	The component of the spectrum-averaged radiation torque
	efficiency vector $\langle\bQ_{\bGam}\rangle$ along $\ahat_1$
	for $a_{\rm eff}=0.2\micron$ grain,
	multiplied by $\bar{\lambda}$,
	as a function of $\cos\Theta$, for the
	five radiation fields of Table \protect{\ref{tab:radfield}}.
	$\langle\bQ_{\bGam}\rangle$ is averaged over
	rotations of the grain around $\ahat_1$.
	}
\end{figure}

\begin{figure}
\epsscale{1.00}
\plotone{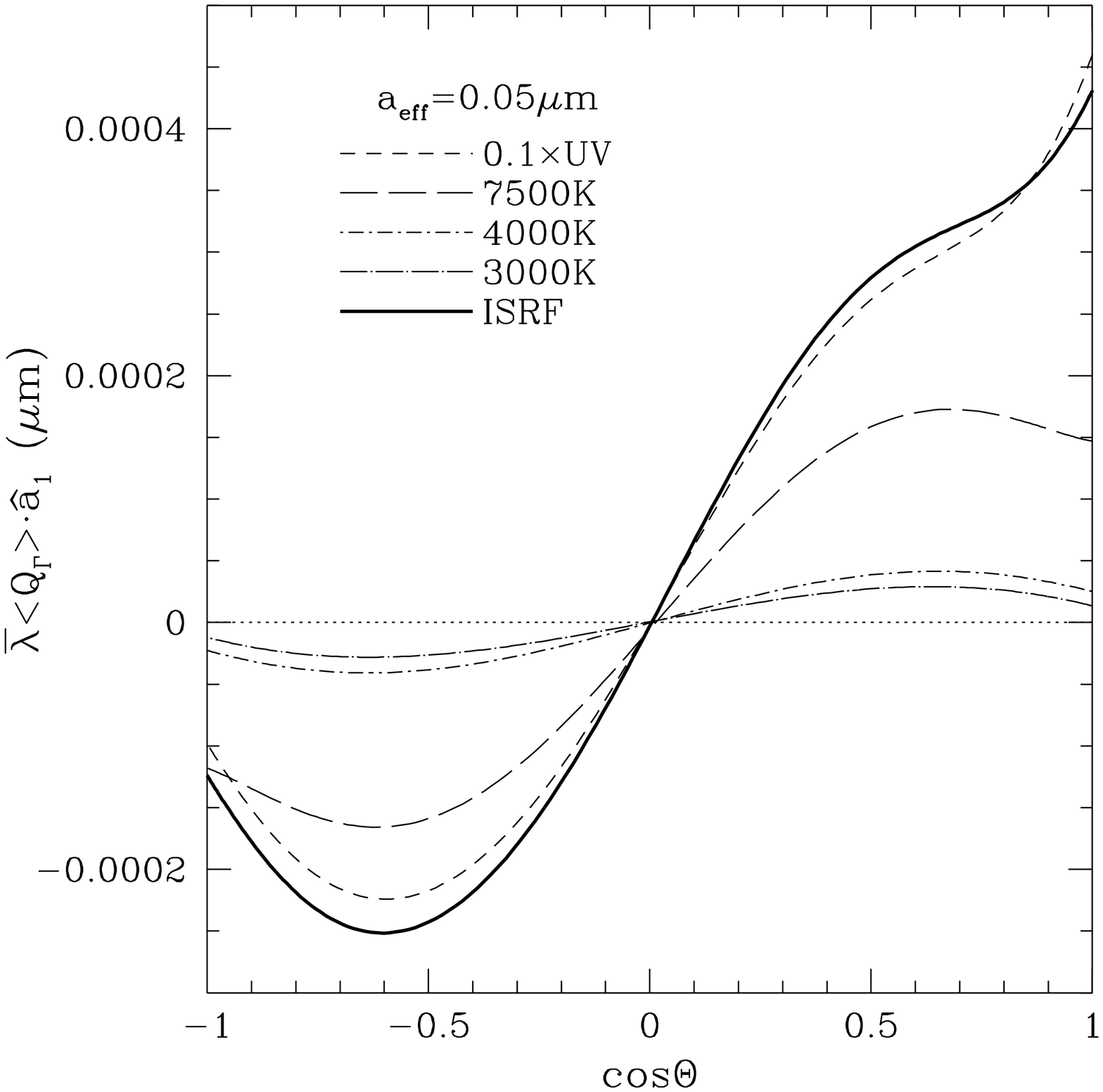}
\caption{
	Same as Fig. \protect{\ref{fig:qgam_a1vsTheta_0.2}}
	but for $a_{\rm eff}=0.05\micron$.
	The results for the ``UV'' radiation field have been
	multiplied by a factor 0.1.
	\label{fig:qgam_a1vsTheta_0.05}
	}
\end{figure}

\begin{figure}
\epsscale{1.00}
\plotone{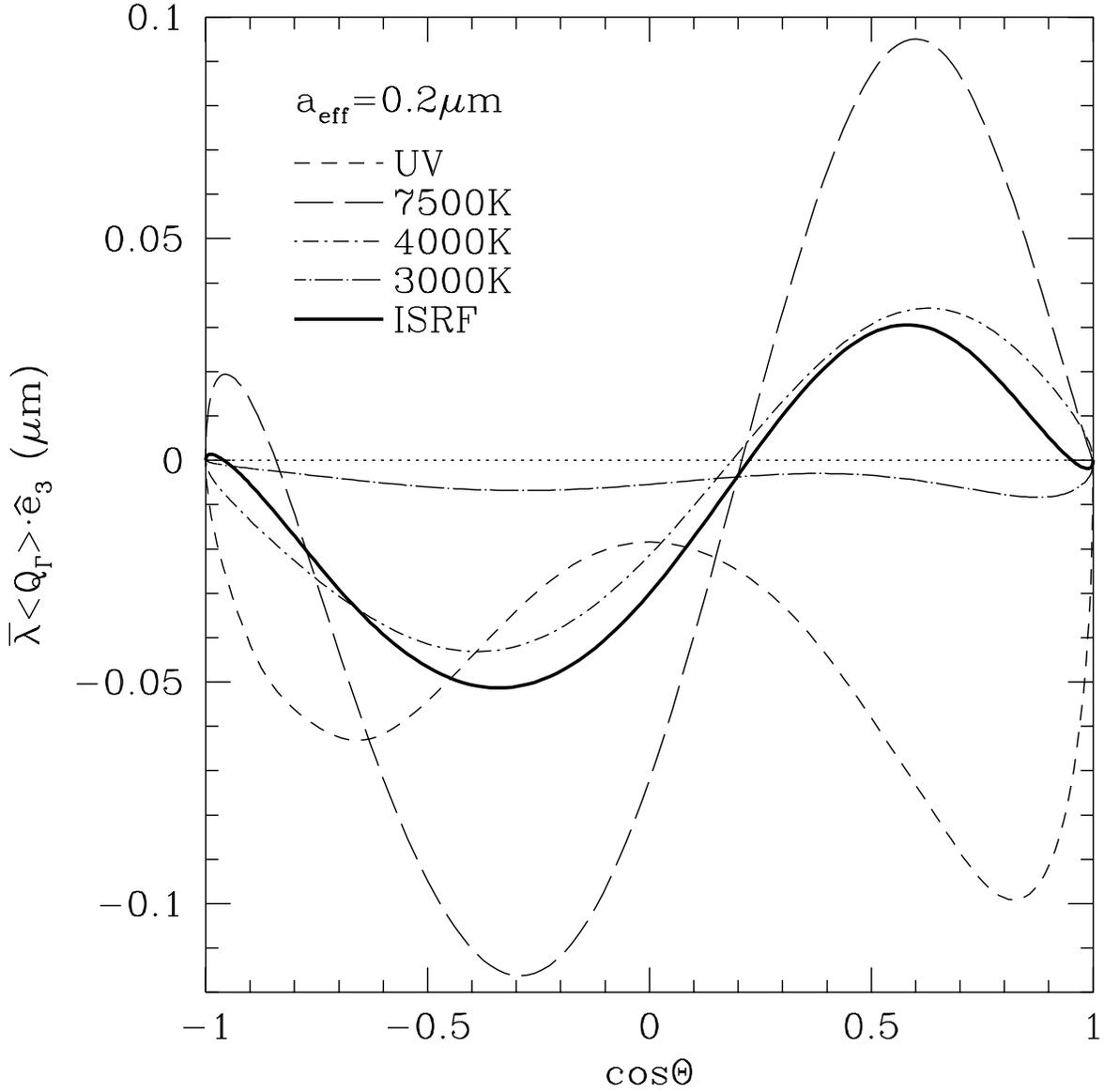}
\caption{
	Same as Fig.\protect{\ref{fig:qgam_a1vsTheta_0.2}} except showing
	the component of the spectrum-averaged radiation torque 
	efficiency vector
	along $\ehat_3$.
	This torque would tend to cause precession of $\bJ$ around
	$\khat$.
	\label{fig:qgam_e3vsTheta_0.2}
	}
\end{figure}

\begin{figure}
\epsscale{1.00}
\plotone{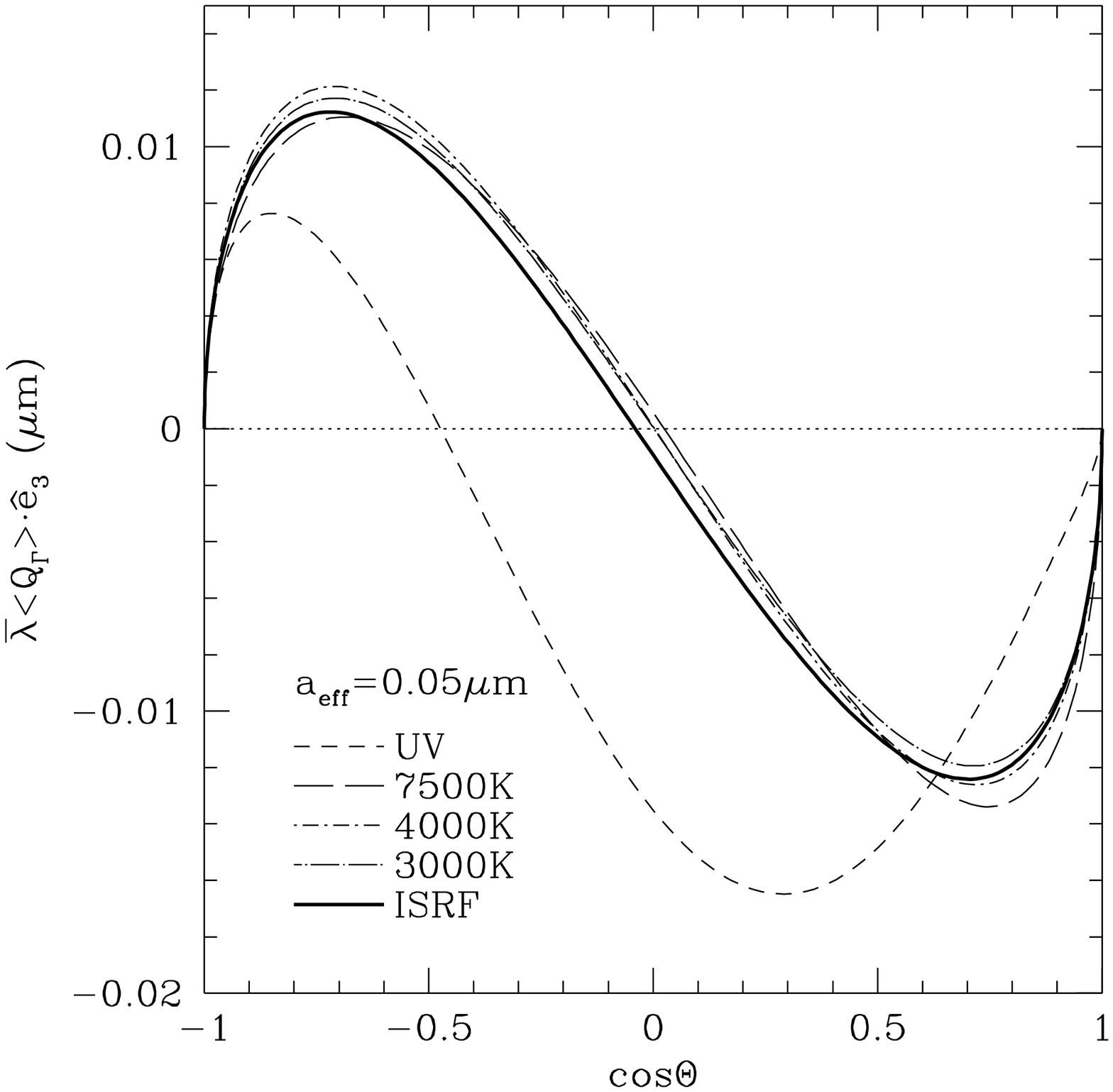}
\caption{
	Same as Fig.\protect{\ref{fig:qgam_e3vsTheta_0.2}}
	except for $a_{\rm eff}=0.05\micron$.
	\label{fig:qgam_e3vsTheta_0.05}
	}
\end{figure}

\begin{figure}
\epsscale{1.00}
\plotone{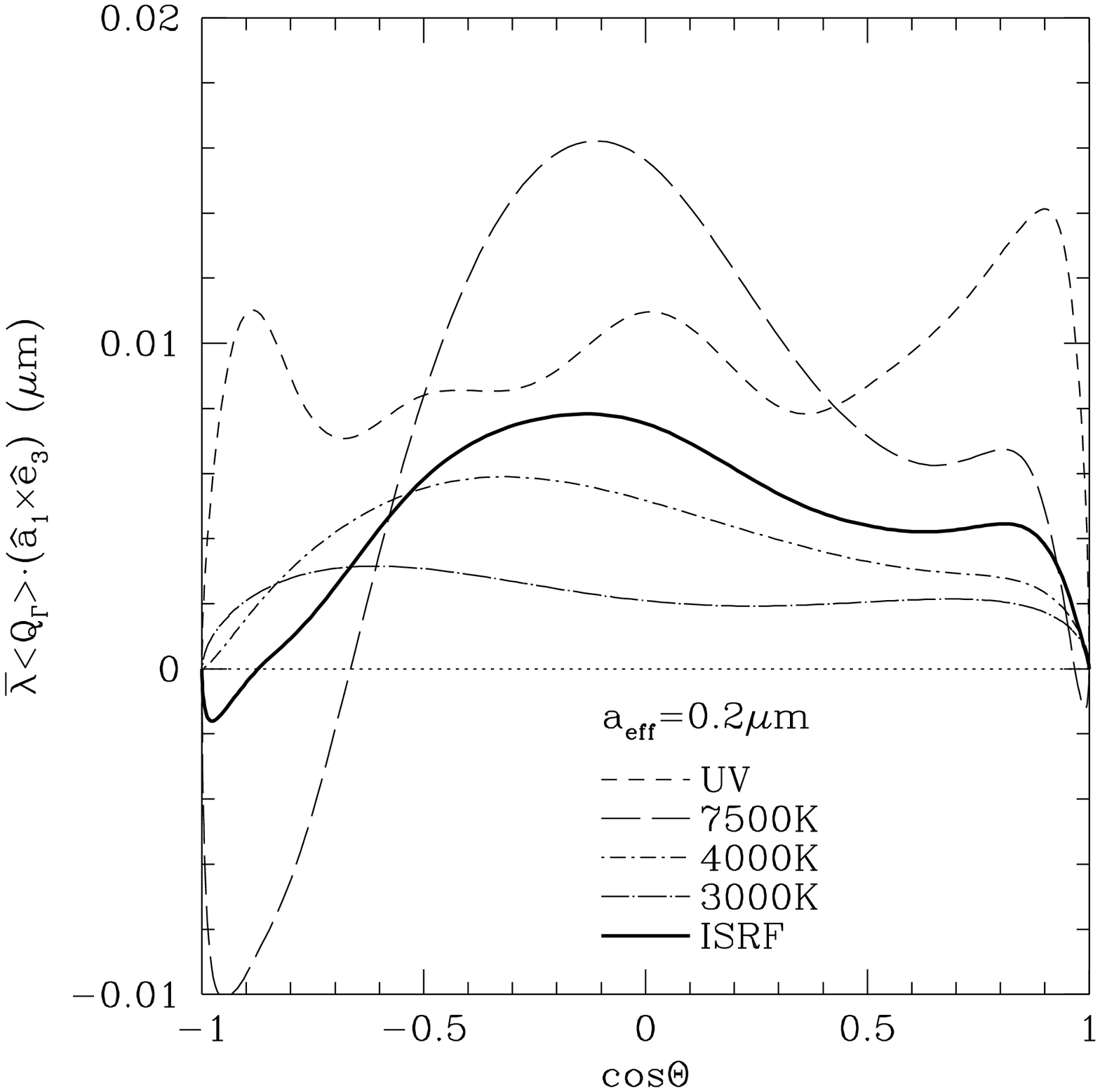}
\caption{
	Same as Fig.\protect{\ref{fig:qgam_a1vsTheta_0.2}} except showing
	the component of the spectrum-averaged torque efficiency vector
	$\langle \bQ_\Gamma\rangle$ along $\ahat_1\times\ehat_3$,
	for $a_{\rm eff}=0.2\micron$ grain.
	If $\ahat_1\parallel\bJ$, this torque (if positive) would tend
	to cause alignment of $\bJ$ with $\khat$.
	\label{fig:qgam_alignvsTheta_0.2}
	}
\end{figure}

\begin{figure}
\epsscale{1.00}
\plotone{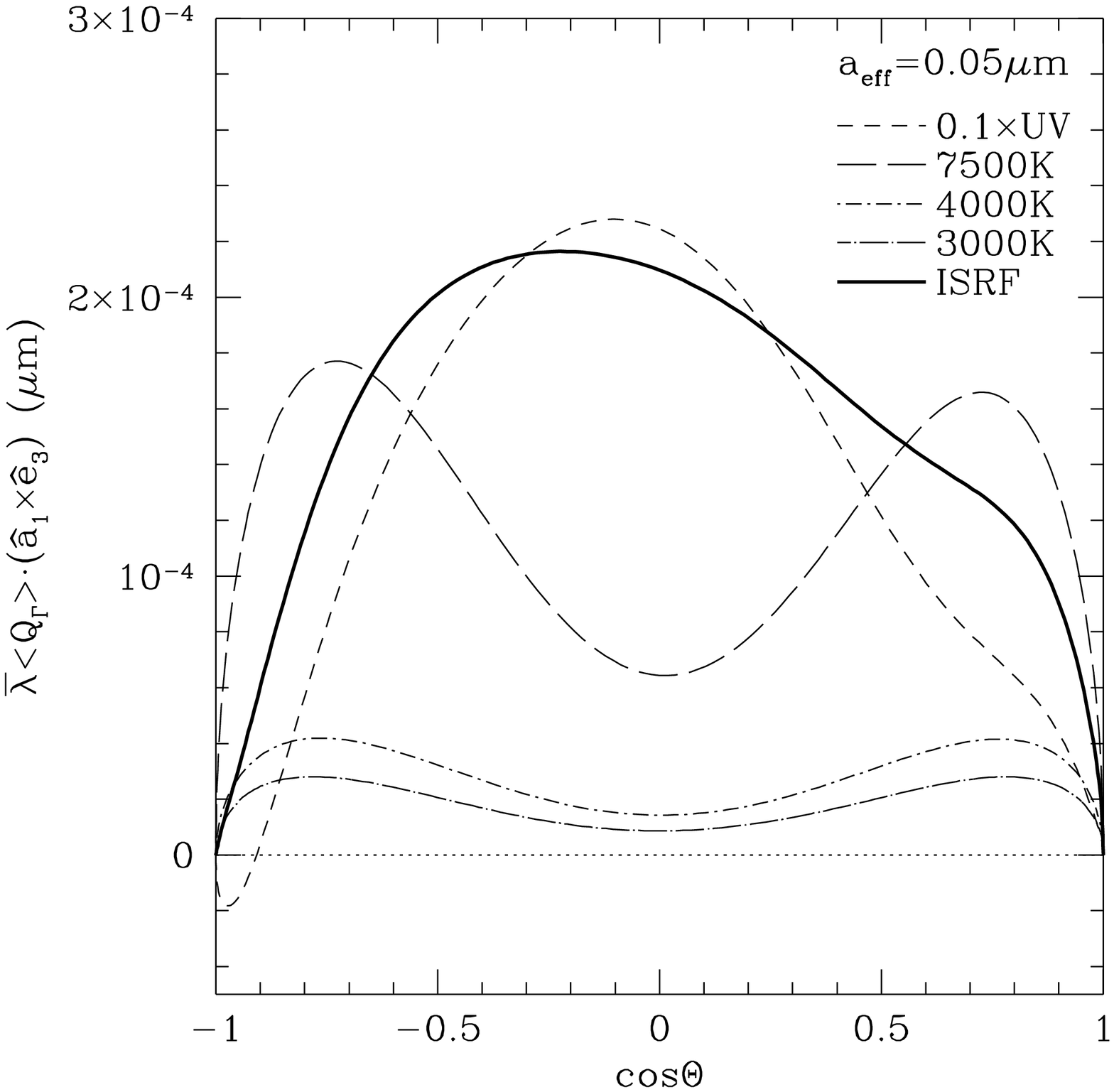}
\caption{
	Same as Fig.\protect{\ref{fig:qgam_alignvsTheta_0.2}}
	but for $a_{\rm eff}=0.05\micron$.
	The results for the ``UV'' radiation field have been
	multiplied by 0.1 .
	\label{fig:qgam_alignvsTheta_0.05}
	}
\end{figure}

In Figs.\ \ref{fig:qgam_a1vsTheta_0.2} -- \ref{fig:qgam_alignvsTheta_0.05} 
we show
the three vector components of 
$\langle\bQ_\Gamma(\Theta,0)\rangle$,
for the $a_\eff=0.2$ and $a_\eff=0.05\micron$ grains; 
results are again shown for each of the $u_\lambda$ in
Table \ref{tab:radfield}.

\subsubsection{Isotropic Radiation}

Consider a grain spinning around axis $\ahat_1$.
An isotropic radiation field
will result in a force $\bF_{rad}$ and torque $\bGam_{rad}$ on the grain which 
are each fixed in body coordinates.
Because of the grain rotation, in inertial coordinates the only
component of the force or torque which will not average to zero will be the
component parallel to the rotation axis $\ahat_1$:
\begin{eqnarray}
\bF_{rad} &=& \pi a_\eff^2(1-\gamma)u_{rad} \langle Q_{pr}^\iso\rangle \ahat_1
\\
\bGam_{rad} &=&
\pi a_\eff^2(1-\gamma)u_{rad}{\lambdabar\over2\pi} 
\langle Q_\Gamma^\iso\rangle \ahat_1
\end{eqnarray}
where
\beq
\langle Q^\iso\rangle \equiv
{1\over 2}\int_0^\pi 
\sin\Theta d\Theta~\ahat_1\!\cdot\!\langle \bQ(\Theta,0)\rangle
\eeq
is the effective efficiency vector for a grain spinning around $\ahat_1$
and exposed to isotropic radiation.
Values of $\langle Q_{pr}^\iso\rangle$ 
and $\langle Q_\Gamma^\iso\rangle$ are given 
in Table \ref{tab:grain_props}.

\subsection{Rotational Damping}

For a grain in neutral gas of hydrogen density $\nH = n(\H)+2n(\HH)$ 
and temperature
$T$,
the gas drag torque on a grain with angular velocity $\omega$ 
around axis $\ahat_1$ may be written
\beq
\bGam_{drag,gas} =
-{2\over 3} \delta \nH (1.2)(8\pi m_\H kT)^{1/2} a_\eff^4 \omega\ahat_1 ~~~.
\label{eq:t_drag}
\eeq
where $m_\H$ is the mass of an H atom.
If all impinging atoms ``stick'' and then leave
with negligible velocity relative to the local (moving) surface,
then $\delta=1$ for a sphere, and
$\delta=2(\pi/3)^{1/3}=2.01$ for a 2:2:1 brick;
in general we expect $\delta\approx\alpha_1$.
The factor 1.2 in eq.~(\ref{eq:t_drag})
allows for the effects of helium with $n_{\He}=0.1\nH$.\footnote{
	We have assumed the H to be fully atomic; in fully molecular gas the
	factor 1.2 in eq.~(\ref{eq:t_drag}) should be 
	replaced by $2^{-1/2}+0.2=.907$.
	}

The rotational damping time is
\beq
\tau_{drag,gas} = {\pi\alpha_1\rho a_\eff \over
3\delta\nH (2\pi m_\H kT)^{1/2}}
= 8.74\times10^4\yr {\alpha_1\over\delta}
\rho_3 a_{-5} T_2^{1/2}
\left({3000\cm^{-3}K\over\nH T}\right)
\eeq
where $\rho_3\equiv\rho/3\g\cm^{-3}$,
$a_{-5}\equiv a_\eff/10^{-5}\cm$,
and $T_2\equiv T/10^2\K$.

In addition to collisions with gas atoms, there is rotational 
damping associated with absorption and emission of
photons by the grain (Purcell \& Spitzer 1971;
Roberge, deGraff \& Flaherty 1993).
If we assume that the grain is heated by starlight to a temperature
$T_d$, and that $Q_{abs}\propto \lambda^{-\beta}$ at wavelengths
$\lambda \gtsim 0.1 hc/kT_d$, then
the damping time due to thermal
emission may be written
\begin{eqnarray}
\tau_{drag,em}&=&{8\alpha_1(\beta+3)\over 5}{\zeta(\beta+4)\over\zeta(\beta+3)}
{\rho a_\eff^3 (kT_d)^2 \over \hbar^2 c u_{rad} \langle Q_{abs} \rangle}
\\
&=& 1.60\times10^5 \yr ~\alpha_1~ \rho_3~ a_{-5}^3 \left({T_d\over 18\K}\right)^2
\left({u_\ISRF\over u_{rad}}\right){1 \over \langle Q_{abs}\rangle}
\label{eq:tau_drag_em}
\end{eqnarray}
where $\zeta(x)$ denotes the Riemann $\zeta$-function,
\beq
\langle Q_{abs}\rangle \equiv
{1\over u_{rad}}\int u_\lambda Q_{abs}(\lambda)d\lambda ~~~,
\eeq
and we have assumed $\beta=2$ in eq.(\ref{eq:tau_drag_em}),
as expected for simple models (Draine \& Lee 1984) and as appears
to be required by the far-infrared emission from dust in diffuse clouds
(Draine 1994).\footnote{
	The damping time $\tau_{drag,em}$ is insensitive to the precise
	value of $\beta$: the factor $(\beta+3)\zeta(\beta+4)/\zeta(\beta+3)$
	decreases by only 22\% if $\beta$ is reduced from $\beta=2$ to 1.
	}
In eq.\ (\ref{eq:tau_drag_em}) it is assumed that the radiated
photons have angular momentum $\hbar$ relative to the grain center of mass;
this will be true for radii $a\ll hc/kT_d = 800 (18\K/T_d)\micron$.
There is additional damping associated with absorption of starlight photons, 
but it
is smaller than the damping due to thermal emission by a factor
$\sim T_d/T_{rad}\approx 1/500$ where $T_{rad}$ is the color temperature
of the radiation responsible for heating the grain (Purcell 1979),
and hence may be neglected.
Values of $\langle Q_{abs}\rangle$ are given in Table \ref{tab:grain_props}.
The rotational damping time $\tau_{drag}$ is given by
\beq
\tau_{drag}^{-1}=\tau_{drag,gas}^{-1}+\tau_{drag,em}^{-1} ~~~.
\label{eq:taudrag}
\eeq

\section{Superthermal Rotation\label{sec:superthermal}}

\subsection{H$_2$ Formation\label{sec:superth_H2}}

The thermal rotation rate for a grain may be written
\beq
\omega_T^2 = {15\over 8\pi\alpha_1} {kT \over\rho a_\eff^5} ~~~,
\eeq
assuming rotation around $\ahat_1$ with kinetic energy $kT/2$.
Purcell (1979) pointed out that $\HH$ formation on the grain
surface, taking place only at randomly-located ``active sites'', would 
result in a large torque.
The component of the torque perpendicular to the rotation axis
$\ahat_1$ would average to zero; for a 2:2:1 rectangular solid
Purcell estimated the component parallel to the
rotation axis
\beq
\bGam_{\HH}\!\cdot\!\ahat_1 = {1\over3}
\left({\pi\over3}\right)^{1/6} f ~n(\H) (2E kT)^{1/2} a_\eff^2 ~l~ p(t)
\eeq
where $n(\H)$ is the density of H atoms in the gas,
$l^2$ is the surface area per $\HH$ formation site on the
grain surface,
$f$ is the fraction of arriving H atoms which depart as $\HH$,
$E$ is the kinetic energy of the departing $\HH$ molecules,
and $p(t)$ is a random variable with time averages
$\langle p(t)\rangle=0$,
$\langle p(t)p(t+\tau)\rangle=e^{-\tau/t_0}$,
where $t_0$ is the ``lifetime'' of a surface recombination site.

If the only torques acting on the grain are those due to gas drag and
$\HH$ formation, the grain will attain a rotation rate 
$\omega_{\HH}$; the mean kinetic energy will exceed $0.5kT$ by a factor
(Purcell 1979)
\beq
\left({\omega_{\HH}\over\omega_T}\right)^2={5\alpha_1 f^2\over 216\,\delta^2}
\left({n(\H)\over\nH}\right)^2 {E\over kT} {\rho a_\eff l^2\over m_\H}
\left({\tau_{drag}\over\tau_{drag,gas}}\right)^2
\left({t_0\over t_0+\tau_{drag}}\right) ~~~.
\eeq

\subsection{Radiative Torques}
If the grain is subject to a steady radiative torque
\beq
\bGam_{rad} = \pi a_\eff^2
u_{rad} {\lambdabar\over 2\pi}
\left[ 
	(1-\gamma)\langle Q_\Gamma^\iso \rangle + 
	\gamma
	\langle \bQ_\Gamma(\Theta)\rangle\!\cdot\!\ahat_1
\right]\ahat_1
\label{eq:gammarad}
\eeq
along the rotation axis $\ahat_1$ then, 
ignoring other sources of rotational excitation by the gas, 
we equate
eqs.\ (\ref{eq:t_drag}) and (\ref{eq:gammarad}) 
to find that the grain will rotate around $\ahat_1$ with an angular velocity
\beq
\omega_{rad} =
{5\bar{\lambda}\over
8 \delta a_\eff^2}
\left({kT\over8\pi m_\H}\right)^{1/2}
\left({u_{rad}\over \nH kT}\right)
\left[ (1-\gamma)\langle Q_\Gamma^\iso\rangle +
	\gamma\langle\bQ_\Gamma \rangle\!\cdot\!\ahat_1
\right]
\left({\tau_{drag}\over\tau_{drag,gas}}\right) ~~~.
\eeq
%1.14\!\times\!10^{10}\s^{-1}
%{T_2^{1/2}\over\delta a_{-5}^2}
%\left( {u_{rad}\over \nH kT}\right)
%\left( {\lambdabar\over\micron} \right)
%\left[(1-\gamma)\langle Q_\Gamma^\iso\rangle +
%	\gamma\langle \bQ_\Gamma\rangle\!\cdot\!\ahat_1
%\right]
%\left({\tau_{drag}\over\tau_{drag,gas}}\right) ~~~.
%\end{eqnarray}
We see that, in addition to $\langle Q_\Gamma^\iso\rangle$ and
$\langle \bQ_\Gamma\rangle\!\cdot\!\ahat_1$,
$\omega_{rad}$ is determined by
the ratio 
of radiation pressure to gas pressure 
$(u_{rad}/\nH kT)$ and the anisotropy factor $\gamma$.
The rotational kinetic energy will exceed $0.5kT$ by a factor
\begin{eqnarray}
\left({\omega_{rad}\over\omega_T}\right)^2
&=&
{5\alpha_1\over 192\delta^2}
\left( {u_{rad} \over \nH k T} \right)^2
\left( \rho a_\eff \lambdabar^2\over m_\H \right)
\left[
	(1-\gamma)\langle Q_\Gamma^\iso\rangle +
	\gamma\langle \bQ_\Gamma \rangle\!\cdot\!\ahat_1
\right]^2
\left({\tau_{drag}\over\tau_{drag,gas}}\right)^2
\\
&=&
4.72\!\times\!10^9 {\alpha_1\over \delta^2}
\rho_3 a_{-5}
\left( {u_{rad}\over \nH kT}\right)^2
\left( {\lambdabar\over\micron} \right)^2
\left[
	(1-\gamma)\langle Q_\Gamma^\iso\rangle +
	\gamma\langle \bQ_\Gamma\rangle \!\cdot\!\ahat_1 
\right]^2
\left({\tau_{drag}\over\tau_{drag,gas}}\right)^2
\label{eq:omegaradoveromegat}
 ~~~.
\end{eqnarray}

\subsection{Diffuse Clouds\label{sec:diffuse}}

In a diffuse cloud where the hydrogen 
is predominantly atomic, $\HH$ recombination
on the grain surface will result in very rapid grain rotation.
Values of $(\omega_{\HH}/\omega_T)^2$ are given in Table \ref{tab:diffuse}
for three grain sizes, and three possible values of the surface recombination
site lifetime $t_0$.

A typical diffuse cloud may have $\nH T= 3000\cm^{-3}\K$;
the average interstellar radiation field (ISRF) then corresponds to
$u_{rad}/\nH kT=2.09$.
We adopt an anisotropy factor $\gamma=0.1$ as representative.

As seen from Figs.\ \ref{fig:qgam_a1vsTheta_0.2} and
\ref{fig:qgam_a1vsTheta_0.05}, the spectrum-averaged
$\langle\bQ_\Gamma(\Theta)\rangle\!\cdot\!\ahat_1$
varies considerably with $\Theta$, the angle between $\ahat_1$
and the radiation flux.
Values of $(\omega_{rad}/\omega_T)^2$ are given in Table \ref{tab:diffuse} 
for $\Theta=0\arcdeg$ and $60\arcdeg$, and
for three grain sizes.

For the $a_\eff=0.2\micron$ grain at $\Theta=0\arcdeg$, we find that radiative
torques result in
extreme superthermal rotation with
$(\omega_{rad}/\omega_T)^2\approx9\times10^4$,
greater than the rotation due to $\HH$ formation on the grain surface
if $l\ltsim35\Angstrom$.
The $a_\eff=0.05\micron$ grain, on the other hand, has a
radiatively-driven rotation rate only slightly in
excess of thermal, with $(\omega_{rad}/\omega_T)^2=18$ 
(for $\Theta=0\arcdeg$), far smaller than the rotation velocity expected
from $\HH$ formation.
The $a_\eff=0.02\micron$ grain is essentially unaffected by radiative
torques: the rotational kinetic energy resulting from these
torques is small compared to $kT$.
We therefore see that --
at least for the particular shape considered here --
radiative torques will drive the larger ($a_\eff\gtsim0.1\micron$) 
grains to highly superthermal
rotation rates, while having only a slight effect on the smaller 
($a_\eff\ltsim0.05\micron$) grains.

It has been known for some time that the observed wavelength dependence of
polarization requires that
if the small grains and large grains have similar shapes, 
then the degree of alignment declines rapidly for $a_\eff\ltsim0.1\micron$
(Mathis 1979, 1986; Kim \& Martin 1994a,b).
This was somewhat unexpected: the standard Davis-Greenstein alignment time
varies as $\tau_{DG}\propto a^2$, so one might naively expect that small
grains would exhibit a {\it higher} degree of alignment than large grains.

Several ideas have been put forward to explain the low degree of
alignment of small grains. 
Mathis (1986) proposed that a grain would not
be aligned unless it contained at least one superparamagnetic inclusion, and 
that small grains were statistically unlikely to contain such an inclusion.
Mathis obtained a good fit to the average polarization curve by assuming that
a grain with $a_\eff=0.08\micron$ had a 50\% probability of containing one
or more inclusions.
Lazarian (1994) argued that if grain alignment was due to the Gold mechanism
(gas-grain streaming) driven by Alfv\'en waves, then only the larger
grains ($a_\eff\gtsim0.1\micron$) were sufficiently inertial to be aligned.
Lazarian (1995c) argued that $\HH$ formation on fractal grains would
only be able to drive grains with $a_\eff\gtsim0.02\micron$ to superthermal
rotation.

As pointed out by Purcell (1975, 1979), superthermal rotation contributes
to grain alignment by rendering unimportant the disaligning effects of
random collisions with gas atoms.
Purcell considered three processes which could drive grains to superthermal
rotation: $\HH$ formation on the grain surface, photoelectric emission,
and variations in the effective accomodation coefficient over the grain
surface;
of the three,
$\HH$ formation appears to be
the most powerful.
As discussed in \S\ref{sec:superth_H2},
the effects of this torque depend on the
``correlation time''  $t_0$ for the torques driving superthermal
rotation.
If the only torques are due to $\HH$ formation and collisions with
gas atoms, the grain will undergo spindown and ``crossover events'' with a
frequency $t_z^{-1}$, where the time $t_z\approx \pi(t_0\tau_{drag})^{1/2}$ 
(Purcell 1979).
If $t_z\ll\tau_{DG}$, then grain alignment by paramagnetic
dissipation will be suppressed,
whereas if $t_z\gg\tau_{DG}$, then paramagnetic dissipation will
result in a high degree of alignment.
The three processes discussed by Purcell depend upon surface properties;
if the grain surface is altered on a relatively short time scale,
then substantial grain alignment will not occur unless the
grains are superparamagnetic (in which case $\tau_{DG}$ is short).

Here we propose another mechanism for selective alignment of the larger
grains.
The radiative torques on the grain depend upon the
global geometry of the grain, and will therefore have a correlation time
of order the grain lifetime, $\sim10^8\yr$.
For the $a_\eff=0.2\micron$ grains considered here, 
$|\bGam_{rad}|^2\gg|\bGam_{\HH}|^2$,
and we may
expect long-lived superthermal rotation.
By itself, long-lived superthermal rotation would lead to
grain alignment by paramagnetic dissipation on the Davis-Greenstein
timescale $\tau_\DG$, as expected for ``Purcell torques'' which are
fixed in body coordinates.
However, when an anisotropic radiation field is present, the
radiative torque components perpendicular to $\ahat_1$
may act directly to change the grain alignment on time scales short
compared to $\tau_\DG$; it appears
that this may be the mechanism responsible for the observed
alignment of $a_\eff\gtsim0.1\micron$ 
dust grains with the interstellar magnetic field
(Draine \& Weingartner 1996).

For the $a_\eff=0.05\micron$ grains, on the other hand, the radiative
torques are weak compared to the Purcell torques due to
$\HH$ formation.
If the surface site lifetime $t_0\ltsim\tau_\DG$, there will be
frequent crossover events, with a resulting low degree of grain
alignment.
This may explain the observed preferential alignment of the larger
$a_\eff\gtsim0.1\micron$ interstellar grains, and the minimal alignment of
$a_\eff\ltsim0.05\micron$ grains.

The dynamics of grain alignment including radiative torques will be
examined in more detail in a later paper (Draine \& Weingartner 1996).

\subsection{Dark Clouds\label{sec:dark}}

The alignment of grains in dense clouds has been problematic.
Recent studies of polarization of background stars 
indicate a low degree of grain alignment
in the central regions of some dark clouds (Goodman et al. 1995).
This 
has been attributed to lack of superthermal rotation in
regions where the hydrogen is predominantly molecular, where the
grains are shielded from ultraviolet photons capable of producing photoelectric
emission, and where the gas and grain temperatures tend to be similar.
However, observations of polarized FIR emission from M17, M42 (Orion), and W3
(Hildebrand et al. 1995; Hildebrand 1996)
shows that the grains responsible for the FIR emission in these dark clouds
are significantly aligned, and it has not been clear why grains in some
dense molecular regions are aligned, while in others they are not.

The dark cloud L1755 observed by Goodman et al. (1995)
has $\nH\approx10^4\cm^{-3}$ and $T\approx20\K$, with a peak extinction through
the cloud $A_V\gtsim 8$.
We suppose the cloud to be externally illuminated by the average
interstellar radiation field.
At the cloud surface we have
$u_{rad}\approx0.5u_\ISRF$ and $\gamma=0.5$, and at
a depth $A\approx2$mag from the cloud surface, we estimate\footnote{
	If the dust has albedo $\sim\!0.5$ and is highly forward-scattering,
	then
	$u_{rad}\approx0.5u_{rad}E_2(0.5A/1.086)$, 
	$\gamma\approx E_3(0.5A/1.086)/E_2(0.5A/1.086)$,
	where $E_2$ and $E_3$ are exponential integrals.
	For $A=2$, we would have
	$u_{rad}\approx.0742u_\ISRF$, 
	$\gamma\approx0.739$.
	}
$u_{rad}\approx0.07u_\ISRF$, $\gamma\approx0.7$; the radiation will of
course be substantially reddened, but we neglect this here.
Table \ref{tab:dark} lists values
of $(\omega_{rad}/\omega_T)^2$ for grains at this depth.
We see that the weakened radiation field, and the higher gas density,
make the radiative torques relatively unimportant in this environment.
Near the cloud surface, of course, conditions are closer to the
diffuse cloud conditions of Table \ref{tab:diffuse}, and radiative
torques will maintain superthermal rotation.
Thus we expect grain alignment near the cloud surface, but not at
depths $A\gtsim 2$mag.

\subsection{Star-Forming Clouds\label{sec:starforming}}

In star-forming clouds, there are obviously sources of starlight within
the clouds.
To estimate the characteristic starlight energy density to which the
FIR-emitting dust is exposed, we note that
for grains with far-infrared emissivity $Q_{abs}\propto \lambda^{-2}$,
the power radiated by the grain $P\propto T_d^6$, where $T_d$ is the
grain temperature.
If we assume that the grain is heated by starlight with an energy density
$u_{rad}$, then,
since the average interstellar radiation field $u_\ISRF$ heats grains to 
$T_d\approx18\K$, we find
\beq
u_{rad}\approx \left( {T_d \over18\K}\right)^6 u_\ISRF ~~~.
\eeq
As an example, we consider one point on the M17 molecular cloud:
$(\alpha,\delta)=(18^h17^m40^s,-15\arcdeg16\arcmin)$.
The $100\micron$ emission at this point has a linear polarization of
$\sim\!5\%$ (Hildebrand et al. 1995), indicating that the grains are
appreciably aligned.
The emission from the dust has a 
$50\micron/100\micron$ color temperature of $\sim85\K$ (Gatley et al. 1979);
for a $\lambda^{-2}$ emissivity this color temperature corresponds to a 
dust temperature $T_d\approx45\K$.
If heated radiatively, these grains must therefore be exposed to a 
radiation field\footnote{
	For $T_d=45\K$, the $50\micron$ optical depth at this position
	is $\tau(50\micron)\approx0.08$, so that the grains must be
	heated by radiation at $\lambda\ll50\micron$.
	}
 with 
$u_{rad}/u_\ISRF\approx (45/18)^6=240$
if the radiation heating the grains has a spectrum similar to 
interstellar starlight (``ISRF'').
While the spectrum no doubt differs from the ISRF spectrum, it is not
obvious how: the embedded massive stars will tend to be bluer, but
the spectrum will be reddened by propagation through the dust.
For the present purposes, we will assume the ISRF spectrum,
with an anisotropy factor $\gamma=0.3$.
The gas in M17 has $\nH\approx 10^5\cm^{-3}$ (cf. Stutzki \& G\"usten 1990),
and presumably has $T\approx45\K$, close to the grain temperature.
We use
eq.(\ref{eq:omegaradoveromegat}) to obtain the
$\omega_{rad}$ values given in Table \ref{tab:m17}.
We see
that $a_\eff=0.2\micron$ grains will have
$(\omega_{rad}/\omega_T)^2\approx 2\times10^4$ (for $\Theta=0\arcdeg$).
Evidently the radiation field within a star-forming region like M17
is sufficiently intense to drive extreme superthermal rotation of the
$a_\eff\gtsim0.1\micron$ grains.
In addition to producing superthermal rotation, the components of
the
radiative torques perpendicular to $\ahat_1$ may
bring about the observed grain alignment
(Draine \& Weingartner 1996).

\section{Summary \label{sec:summary}}

The principal results in this paper are as follows:
\begin{enumerate}
\item We show how forces and torques on an irregular grain may be calculated
using the discrete dipole approximation.
\item We report numerical results 
(in Fig.\ \ref{fig:qpr_kvslambda} and Table \ref{tab:grain_props})
for $\khat\!\cdot\!\bQ_{pr}(\lambda)$,
the component of the radiation pressure efficiency vector parallel
to the radiative flux
for one particular irregular grain geometry,
where the grain is assumed to be spinning around its principal
axis $\ahat_1$ of largest moment of inertia.
\item We report numerical results for the
``radiation torque efficiency vector'' $\bQ_\Gamma(\lambda)$,
(in Figs.\ \ref{fig:qgam_a1vslambda_0.2} -- \ref{fig:qgam_alignvsTheta_0.05})
for one irregular grain geometry.
The torque 
depends upon the angle $\Theta$ between the incident flux and the grain
rotation axis $\ahat_1$.
\item In interstellar diffuse clouds, 
radiative torques will drive $a_\eff\gtsim0.1\micron$ grains 
to extreme superthermal rotation.
\item Radiative torques will dominate the torques due
to $\HH$ formation for $a_\eff\gtsim0.1\micron$ interstellar grains.
For these grains, the superthermal rotation is expected to be very
long-lived, since it depends upon global
properties of the grain, rather than the relatively short-lived
surface properties which determine superthermal rotation 
due to $\HH$ formation,
photoelectric emission,
or variations in surface accomodation coefficient.
\item In an isotropic radiation field, the long-lived superthermal 
rotation of $a_\eff\gtsim0.1\micron$ grains due to radiative torques
would facilitate alignment of these grains with the Galactic
magnetic field by the Davis-Greenstein mechanism of paramagnetic relaxation.
\item Since the rotation of smaller ($a\ltsim0.05\micron$) grains is
not dominated by radiative torques, alignment of these grains with
the Galactic magnetic field would be limited due to random variations in the
torque associated with $\HH$ formation on the grain surface.
This may explain why $a_\eff\gtsim0.1\micron$ grains in diffuse clouds
are aligned, while there is evidently minimal alignment of the
$a_\eff\ltsim0.05\micron$ grains which dominate the extreme ultraviolet
extinction.
\item Radiative torques appear to be able to drive grains to superthermal
rotation in star-forming regions, such as M17, where the ratio of
anisotropic radiation to gas pressure is relatively high, thereby
potentially explaining the observed alignment of far-infrared emitting
dust in M17 and other star-forming regions with warm dust.
In cold dark clouds, on the other hand, radiative torques are unable
to drive the grains to superthermal rotation, consistent with the
observed lack of aligned grains deep in quiescent dark clouds.
\item In addition to the torque component parallel to the grain 
rotation axis $\ahat_1$
(which drives superthermal rotation), anisotropy of the radiation field
can result in radiative torque
components perpendicular to the grain angular momentum; 
these components can affect grain alignment directly.
\end{enumerate}

\acknowledgements
This research was supported in part by NSF grant
AST-9219283 to BTD, and by an NSF Graduate Research Fellowship
to JCW.
We are grateful to R.H. Lupton for the availability of the SM plotting package,
and to A. Lazarian, W.G. Roberge, and L. Spitzer for helpful comments.
\appendix
\section{Appendix: Radiation Field due to Oscillating Dipoles}

Consider an array of oscillating 
point electric dipoles $\bp_j$, at locations $\br_j$.
At large distances $r$ 
the fields due to these dipoles may be written
\begin{eqnarray}
\bE_{sca} &=&
e^{ikr}
\sum_{j=1}^N
\biggl\{
{k^2\over r}\left[\bp_j - \nhat(\nhat\!\cdot\!\bp_j)\right]
+
\bp_j{k^2\over r^2}
	\left[
	(\nhat\!\cdot\!{\br_j}) + 
	{i k(r_j^2-(\nhat\!\cdot\!\br_j)^2)\over 2} +
	{i\over k}
	\right]
\nonumber
\\
&~&~~~~~~~~~~+~
\nhat{k^2\over r^2}
	\left[
	(\br_j\!\cdot\!\bp_j) -
	(\nhat\!\cdot\!\bp_j)
		\left(
		3(\nhat\!\cdot\!\br_j) +
		{3i\over k} +
		{ik\over 2}
			\left[ r_j^2-(\nhat\!\cdot\!\br_j)^2 \right]
		\right)
	\right]
\nonumber
\\
&~&~~~~~~~~~~+~
\br_j{k^2\over r^2}(\nhat\!\cdot\!\bp_j)
\biggr\}
\exp(-ik\nhat\!\cdot\!\br_j)~+~ O(r^{-3})
\\
\bB_{sca} &=&
e^{ikr}
\sum_{j=1}^N
\biggl\{
{k^2\over r}\nhat\times\bp_j
+
{k^2\over r^2}
	\left[
	2(\nhat\!\cdot\!\br_j) +
	{ik\over2}(r_j^2-(\nhat\!\cdot\!\br_j)^2) -
	{1\over ik}
	\right]
\nhat\times\bp_j
\nonumber
\\
&~&~~~~~~~~~~-~
{k^2\over r^2}\br_j\times\bp_j
\biggr\}
\exp(-ik\nhat\!\cdot\!\br_j)~+~ O(r^{-3})
\end{eqnarray}
Thus,
\begin{equation}
\nhat\!\cdot\!\bE_{sca} = {k^2\over r^2}
e^{ikr}
\sum_{j=1}^N \left\{ (\br_j\!\cdot\!\bp_j) - 
(\nhat\!\cdot\!\bp_j) \left[(\nhat\!\cdot\!\br_j)+{2i\over k}\right]
\right\}\exp(-ik\nhat\!\cdot\!\br_j)~+~ O(r^{-3})
\end{equation}
\begin{equation}
\bE_{sca}\times\nhat = 
{k^2\over r}
e^{ikr}
\sum_{j=1}^N \bp_j\times\nhat
\exp(-ik\nhat\!\cdot\!\br_j)~+~ O(r^{-2})
\end{equation}
\begin{equation}
\nhat\!\cdot\!\bB_{sca} = {k^2\over r^2}
e^{ikr}
\nhat\!\cdot\!\sum_{j=1}^N \bp_j\times\br_j
\exp(-ik\nhat\!\cdot\!\br_j)~+~ O(r^{-3})
\end{equation}
\begin{equation}
\bB_{sca}\times\nhat = {k^2\over r}
e^{ikr}
\sum_{j=1}^N 
\left[\bp_j - \nhat(\nhat\!\cdot\!\bp_j)\right]
\exp(-ik\nhat\!\cdot\!\br_j)~+~ O(r^{-2})
\end{equation}
These expressions, substituted into eq.(\ref{eq:gamout}),
lead to the result (\ref{eq:gamout2}) for the time-averaged
torque $\langle\bGam_{out}\rangle$.

%************************* table 1 *****************************************
\begin{deluxetable}{c c c c }
\tablecaption{Target Geometry: Coordinates of Constituent Blocks
	\label{tab:cubes}}

\tablehead{
\colhead{$j$}
&\colhead{$x_j$}
&\colhead{$y_j$}
&\colhead{$z_j$}
	}
\startdata
1&	0&	1&	0\nl
2&	0&	1&	1\nl
3&	0&	2&	0\nl
4&	0&	2&	1\nl
5&	1&	1&	0\nl
6&	1&	1&	1\nl
7&	1&	2&	0\nl
8&	1&	2&	1\nl

9&	0&	0&	1\nl
10&	0&	0&	2\nl
11&	0&	1&	2\nl
12&	2&	1&	0\nl
13&	2&	2&	0\nl

\enddata
\end{deluxetable}
%************************ end table 1 ***************************************
%************************* table 2 *****************************************
\begin{deluxetable}{c c c c c}
\tablecaption{Target Principal Axes $\ahat_j$ and factors $\alpha_j$
	\label{tab:axes}}

\tablehead{
\colhead{$j$}
&\colhead{$(\ahat_j)_x$}
&\colhead{$(\ahat_j)_y$}
&\colhead{$(\ahat_j)_z$}
&\colhead{$\alpha_j$}
	}
\startdata
1	&0.4488	& 0.4357	& 0.7802&	1.745\nl
2	&0.6673	&-0.7441	& 0.0317&	1.610\nl
3	&0.5944	& 0.5064	&-0.6247&	0.876\nl
\enddata
\end{deluxetable}
%************************ end table 2 ***************************************
%********************* table 3 **********************************************
\begin{deluxetable}{c c c}
\tablecaption{Interstellar Radiation Field Components\label{tab:radfield}}
\tablehead{
\colhead{Radiation Field}
&\colhead{$u_{rad}$}
&\colhead{$\bar{\lambda}$}
%&\colhead{$\langle Q_\Gamma\rangle$}
\\
\colhead{}
&\colhead{$\erg\cm^{-3}$}
&\colhead{$\micron$}
	}
\startdata
UV [eq.(\ref{eq:uvfield})]
	&$7.13\times10^{-14}$
		&0.1566\nl
$W=1\times10^{-14}$, $T=7500\K$  ($\lambda>2460\Angstrom$)
	&$2.29\times10^{-13}$
		&0.7333\nl
$W=1.65\times10^{-13}$, $T=4000\K$ ($\lambda>2460\Angstrom$)
	&$3.19\times10^{-13}$
		&1.3319\nl
$W=4\times10^{-13}$, $T=3000\K$ ($\lambda>2460\Angstrom$)
	&$2.45\times10^{-13}$
		&1.7755\nl
ISRF
	&$8.64\times10^{-13}$
		&1.2021\nl
\enddata
\end{deluxetable}
%***************************end table 3************************************
%**************************** table 4 *************************************
\begin{deluxetable}{l c c c}
\tablecaption{Silicate Grain Properties\label{tab:grain_props}}
\tablehead{
\colhead{}
&\colhead{$a_\eff=0.02\micron$}
&\colhead{$a_\eff=0.05\micron$}
&\colhead{$a_\eff=0.2\micron$}
	}
\startdata
$\langle Q_{abs}\rangle_{uv}$
	&0.758			&1.254			&1.100\nl
$\langle Q_{abs}\rangle_{7500\K}$
	&0.0150			&0.0479			&0.340\nl
$\langle Q_{abs}\rangle_{4000\K}$
	&0.00855			&0.0234			&0.163\nl
$\langle Q_{abs}\rangle_{3000\K}$
	&0.00664			&0.0181			&0.111\nl
$\langle Q_{abs}\rangle_\ISRF$
	&0.0716			&0.130			&0.273\nl
\hline
$\langle Q_{pr}^\iso\rangle_{uv}$
	&$1.31\times10^{-3}$	&$-6.82\times10^{-3}$	&$-4.12\times10^{-3}$\nl
$\langle Q_{pr}^\iso\rangle_{7500}$
	&$2.\times10^{-7}$	&$1.29\times10^{-4}$	&$-2.25\times10^{-2}$\nl
$\langle Q_{pr}^\iso\rangle_{4000}$
	&--			&$5.4\times10^{-6}$	&$3.76\times10^{-3}$\nl
$\langle Q_{pr}^\iso\rangle_{3000}$
	&--			&--			&$-6.53\times10^{-4}$\nl
$\langle Q_{pr}^\iso\rangle_\ISRF$
	&$1.1\times10^{-4}$	&$-5.25\times10^{-4}$	&$-7.86\times10^{-3}$\nl
\hline
$\langle Q_\Gamma^\iso\rangle_{uv}$
	&$2.64\times10^{-4}$	&$2.52\times10^{-3}$		&$5.34\times10^{-3}$\nl
$\langle Q_\Gamma^\iso\rangle_{7500\K}$
	&$4.43\times10^{-8}$	&$2.04\times10^{-6}$		&$5.31\times10^{-4}$\nl
$\langle Q_\Gamma^\iso\rangle_{4000\K}$
	&$1.41\times10^{-8}$	&$1.89\times10^{-7}$		&$1.06\times10^{-4}$\nl
$\langle Q_\Gamma^\iso\rangle_{3000\K}$
	&$9.03\times10^{-9}$	&$6.48\times10^{-8}$		&$2.99\times10^{-5}$\nl
$\langle Q_\Gamma^\iso\rangle_\ISRF$
	&$2.84\times10^{-6}$	&$2.75\times10^{-5}$		&$1.99\times10^{-4}$\nl
\hline
$\khat\!\cdot\!\langle \bQ_{pr}\rangle_{uv}$   ($\Theta=0\arcdeg$)
	&1.141			&1.860			&1.438\nl
$\khat\!\cdot\!\langle \bQ_{pr}\rangle_{7500\K}$   ($\Theta=0\arcdeg$)
	&0.0203			&0.184			&1.117\nl
$\khat\!\cdot\!\langle \bQ_{pr}\rangle_{4000\K}$   ($\Theta=0\arcdeg$)
	&0.00965		&0.0454			&0.651\nl
$\khat\!\cdot\!\langle \bQ_{pr}\rangle_{3000\K}$   ($\Theta=0\arcdeg$)
	&0.00724		&0.0257			&0.457\nl
$\khat\!\cdot\!\langle \bQ_{pr}\rangle_\ISRF$   ($\Theta=0\arcdeg$)
	&0.105			&0.226			&0.784\nl
\hline
$\ahat_1\!\cdot\!\langle\bQ_\Gamma\rangle_{uv}$  ($\Theta=0\arcdeg$)
	&$2.54\times10^{-3}$	&$2.94\times10^{-2}$	&$7.01\times10^{-2}$\nl
$\ahat_1\!\cdot\!\langle\bQ_\Gamma\rangle_{uv}$ ($\Theta=60\arcdeg$)
	&$3.55\times10^{-3}$	&$1.67\times10^{-2}$	&$9.06\times10^{-2}$\nl
$\ahat_1\!\cdot\!\langle\bQ_\Gamma\rangle_{7000\K}$  ($\Theta=0\arcdeg$)
	&$6.01\times10^{-6}$	&$2.00\times10^{-4}$	&$7.45\times10^{-2}$\nl
$\ahat_1\!\cdot\!\langle\bQ_\Gamma\rangle_{7000\K}$ ($\Theta=60\arcdeg$)
	&$1.22\times10^{-5}$	&$2.16\times10^{-4}$	&$4.38\times10^{-3}$\nl
$\ahat_1\!\cdot\!\langle\bQ_\Gamma\rangle_{4000\K}$  ($\Theta=0\arcdeg$)
	&$1.60\times10^{-6}$	&$1.88\times10^{-5}$	&$7.78\times10^{-3}$\nl
$\ahat_1\!\cdot\!\langle\bQ_\Gamma\rangle_{4000\K}$ ($\Theta=60\arcdeg$)
	&$3.88\times10^{-6}$	&$2.90\times10^{-5}$	&$4.39\times10^{-3}$\nl
$\ahat_1\!\cdot\!\langle\bQ_\Gamma\rangle_{3000\K}$  ($\Theta=0\arcdeg$)
	&$9.01\times10^{-7}$	&$7.36\times10^{-6}$	&$2.07\times10^{-3}$\nl
$\ahat_1\!\cdot\!\langle\bQ_\Gamma\rangle_{3000\K}$ ($\Theta=60\arcdeg$)
	&$2.27\times10^{-6}$	&$1.52\times10^{-5}$	&$2.07\times10^{-3}$\nl
$\ahat_1\!\cdot\!\langle\bQ_\Gamma\rangle_\ISRF$  ($\Theta=0\arcdeg$)
	&$2.93\times10^{-5}$	&$3.59\times10^{-4}$	&$1.69\times10^{-2}$\nl
$\ahat_1\!\cdot\!\langle\bQ_\Gamma\rangle_\ISRF$ ($\Theta=60\arcdeg$)
	&$4.26\times10^{-5}$	&$2.33\times10^{-4}$	&$4.35\times10^{-3}$\nl
\enddata
\end{deluxetable}
%************************* end table 4 ***********************************
%******************** table 5 *******************************************
\begin{deluxetable}{c c c c}
\tablecaption{Silicate Grains in Diffuse Clouds\label{tab:diffuse}}
\tablehead{
\colhead{}
&\colhead{$a_\eff=0.02\micron$}
&\colhead{$a_\eff=0.05\micron$}
&\colhead{$a_\eff=0.2\micron$}
	}
\startdata
$\tau_{drag,gas}$ (yr)
	&$1.53\times10^4$	&$3.81\times10^4$	&$1.53\times10^5$\nl
$\tau_{drag,em}$ (yr)
	&$3.12\times10^4$	&$2.68\times10^5$	&$8.18\times10^6$\nl
$\tau_{drag}$ (yr)
	&$1.03\times10^4$	&$3.34\times10^4$	&$1.50\times10^5$\nl
$\omega_T$ (rad $\s^{-1}$)
	&$7.01\times10^6$	&$7.10\times10^5$	&$2.22\times10^4$\nl
$(\omega_{\H2}/\omega_T)^2$ for $t_0=10^4\yr$
	&$2.10\times10^2$	&$4.16\times10^2$	&$5.65\times10^2$\nl
$(\omega_{\H2}/\omega_T)^2$ for $t_0=10^5\yr$
	&$3.86\times10^2$	&$1.35\times10^3$	&$3.62\times10^3$\nl
$(\omega_{\H2}/\omega_T)^2$ for $t_0=10^6\yr$
	&$4.22\times10^2$	&$1.75\times10^3$	&$7.86\times10^3$\nl
$(\omega_{rad}/\omega_T)^2$ for $\Theta=0\arcdeg$
	&$3.56\times10^{-2}$	&$18.4$			&$8.73\times10^4$	\nl
$(\omega_{rad}/\omega_T)^2$ for $\Theta=60\arcdeg$
	&$5.49\times10^{-2}$	&11.5			&$1.00\times10^4$	\nl
\enddata
\tablecomments{
	We assume the grain properties of 
	Table \protect{\ref{tab:grain_props}},
	$\rho=3\g\cm^{-3}$,
	$\alpha_1=1.745$, 
	$\delta=2$,
	$\nH=30\cm^{-3}$,
	$T=100\K$,
	$T_d=18\K$,
	$f=1/3$, 
	$n(\H)/\nH=1$,
	$E_{\HH}=0.2\eV$, 
	$l=10\Angstrom$, 
	$u_{rad}=u_\ISRF$,
	and
	$\gamma=0.1$.
	}
\end{deluxetable}
%***************************** end table 5 ********************************
\clearpage
%******************** table 6 *******************************************
\begin{deluxetable}{c c c c}
\tablecaption{Silicate Grains in Dark Clouds\label{tab:dark}}
\tablehead{
\colhead{}
&\colhead{$a_\eff=0.02\micron$}
&\colhead{$a_\eff=0.05\micron$}
&\colhead{$a_\eff=0.2\micron$}
	}
\startdata
$\tau_{drag,gas}$ (yr)
	&$1.02\times10^2$	&$2.56\times10^2$	&$1.02\times10^3$\nl
$\tau_{drag,em}$ (yr)
	&$9.63\times10^5$	&$8.29\times10^6$	&$2.52\times10^8$\nl
$\tau_{drag}$ (yr)
	&$1.02\times10^2$	&$2.56\times10^2$	&$1.02\times10^3$\nl
$\omega_T$ (rad $\s^{-1}$)
	&$3.14\times10^6$	&$3.17\times10^5$	&$9.92\times10^3$\nl
$(\omega_{\H2}/\omega_T)^2$ for $t_0=10^4\yr$
	&$0.466$		&$1.15$			&$4.27$\nl
$(\omega_{\H2}/\omega_T)^2$ for $t_0\gtsim10^5\yr$
	&$0.470$		&$1.17$			&$4.70$\nl
$(\omega_{rad}/\omega_T)^2$ for $\Theta=0\arcdeg$
	&$1.30\times10^{-6}$	&$9.63\times10^{-4}$	&$4.04$\nl
$(\omega_{rad}/\omega_T)^2$ for $\Theta=60\arcdeg$
	&$5.39\times10^{-6}$	&$2.10\times10^{-4}$	&$0.275$\nl
\enddata
\tablecomments{
	We assume the grain parameters of Table
	\protect{\ref{tab:grain_props}},
	$\rho=3\g\cm^{-3}$,
	$\alpha_1=1.745$, 
	$\delta=2$,
	$\nH=10^4\cm^{-3}$,
	$T=20\K$,
	$f=1/3$, 
	$n(\H)/\nH=.01$,
	$E_{\HH}=0.2\eV$, 
	$l=10\Angstrom$, 
	$T_d=10\K$,
	$u_{rad}=.07u_\ISRF$,
	and
	$\gamma=0.7$.
	}
\end{deluxetable}
%***************************** end table 6 ********************************
%******************** table 7 *******************************************
\begin{deluxetable}{c c c c}
\tablecaption{Silicate Grains in Star-Forming Clouds\label{tab:m17}}
\tablehead{
\colhead{}
&\colhead{$a_\eff=0.02\micron$}
&\colhead{$a_\eff=0.05\micron$}
&\colhead{$a_\eff=0.2\micron$}
	}
\startdata
$\tau_{drag,gas}$ (yr)
	&$6.82$			&$17.1$			&$68.2$\nl
$\tau_{drag,em}$ (yr)
	&$812.$			&$6.99\times10^3$	&$2.13\times10^5$\nl
$\tau_{drag}$ (yr)
	&$6.76$			&$17.1$			&$68.2$\nl
$\omega_T$ (rad $\s^{-1}$)
	&$4.70\times10^6$	&$4.76\times10^5$	&$1.49\times10^4$\nl
$(\omega_{\H2}/\omega_T)^2$ for $t_0\gtsim10^4\yr$
	&$0.209$		&$0.522$		&$2.09$\nl
$(\omega_{rad}/\omega_T)^2$ for $\Theta=0\arcdeg$
	&$7.59\times10^{-3}$	&$2.65$			&$1.79\times10^4$\nl
$(\omega_{rad}/\omega_T)^2$ for $\Theta=60\arcdeg$
	&$1.42\times10^{-2}$	&$1.31$			&$1.38\times10^3$\nl
\enddata
\tablecomments{
	We assume the grain parameters of Table 
	\protect{\ref{tab:grain_props}},
	$\rho=3\g\cm^{-3}$,
	$\alpha_1=1.745$, 
	$\delta=2$,
	$\nH=10^5\cm^{-3}$,
	$T=45\K$,
	$f=1/3$, 
	$n(\H)/\nH=.01$,
	$E_{\HH}=0.2\eV$, 
	$l=10\Angstrom$, 
	$T_d=45\K$,
	$u_{rad}=240u_\ISRF$,
	and
	$\gamma=0.3$.
	}
\end{deluxetable}
%***************************** end table 7 ********************************
%                            begin figures
%\clearpage	%seems to be required to avoid overflow in latex...
\end{document}